\documentclass[aps,12pt]{article}
\usepackage{amsfonts}
\usepackage{amssymb}
\usepackage{amsmath}
\usepackage{graphicx}

\usepackage{anysize}
\usepackage{theorem}
\usepackage{latexsym}
\usepackage{eufrak}
\usepackage{syntonly}
\usepackage[mathcal]{euscript}
\usepackage{pb-diagram}
\usepackage{epsf}

\usepackage[latin1]{inputenc}
\usepackage[T1]{fontenc}

\newcommand{\be}{\begin{equation}}
\newcommand{\ee}{\end{equation}}
\newcommand{\ba}{\begin{eqnarray}}
\newcommand{\ea}{\end{eqnarray}}

\input epsf

\def\a{\alpha}
\def\ah{\hat\a}

\def\au{\underline{a}}
\def\b{\beta}

\def\bh{\hat\b}
\def\bu{\underline{b}}
\def\cu{\underline{c}}
\def\d{\delta}

\def\du{\underline{d}}
\def\dh{\hat\d}
\def\dt{\widetilde d}

\def\e{\epsilon}

\def\eu{\underline{e}}
\def\eh{\hat\e}

\def\fu{\underline{f}}
\def\g{\gamma}

\def\gu{\underline{g}}
\def\gh{\hat\g}
\def\hu{\underline{h}}

\def\l{\lambda}

\def\lh{\widehat\lambda}

\def\o{\omega}

\def\oh{\widehat\o}

\def\O{\Omega}

\def\Ot{\widetilde\O}

\def\t{\theta}

\def\G{\Gamma}

\def\Nh{\hat N}
\def\N{\nabla}
\def\Nb{\overline\nabla}

\def\p{\partial}
\def\pb{\overline\partial}
\def\Jb{\overline J}
\def\Ct{\widetilde C}

\def\St{\widetilde S}

\begin{document}

\begin{flushright}
\end{flushright}

\vskip 0.7in
\center{{\LARGE Quantum Current Algebra for the  $AdS_5 \times S^5$ Superstring}}

\vspace{25pt}

{\large Oscar A. Bedoya $^a $\footnote{e-mail:
abedoya@fma.if.usp.br}, Dafni Z. Marchioro $^b $\footnote{e-mail:
dafnimarchioro@unipampa.edu.br},\ Daniel L. Nedel
$^b$\footnote{e-mail: daniel.nedel@unipampa.edu.br} \\and  Brenno
Carlini Vallilo$^c$\footnote{e-mail: vallilo@unab.cl } }

\vspace{15pt}

\center{


$^a${\em Instituto de F\'{\i}sica,
Universidade de S\~ao Paulo}, \\ 05314-970, S\~ao Paulo, SP, 
Brasil. }
\vskip 0.1in
$^b${\em Universidade Federal do Pampa - UNIPAMPA \\
Rua Carlos Barbosa s/n, Bairro Get\'ulio Vargas, CEP 96412-420, Bag\'e - RS, 
Brazil}
\vskip 0.1in
$^c${\em Departamento de Ciencias F\'\i sicas,
Universidad Andres Bello \\ Republica 220, Santiago, Chile} 

\vspace{9pt}

\abstract{The sigma model describing the dynamics of the superstring  in the $AdS_5 \times S^5$ background can be constructed using the coset $PSU(2,2|4)/SO(4,1)\times SO(5)$. A basic set of operators in this two dimensional conformal field theory is composed by the left invariant currents.  Since these currents are not (anti) holomorphic, their OPE's is not determined by
symmetry principles and its computation should be performed
perturbatively. Using the pure spinor sigma model for this background,
we compute the one-loop correction to these OPE's. We also compute the OPE's of the left invariant currents with the energy momentum tensor at tree level and one loop. }


\newpage

\section{Introduction} 
During the last few years a number of results were obtained in $N=4$
super Yang-Mills using integrability techniques\footnote{The
literature on this subject is very large, and we did not attempt to
give a list of references.} culminating in a general system of equations 
that predicts the anomalous dimension for all operators at any value of the 
coupling constant \cite{gromov} . Despite all these results, the quantum 
properties of the dual string theory, namely strings in $AdS_5\times S^5$, 
still remain elusive. 

The classical integrability of the $AdS_5\times S^5$ sigma model was 
established in the paper \cite{Bena:2003wd} and later is was shown to hold 
also in the pure spinor description \cite{Vallilo:2003nx}. Using 
cohomological and algebraic renormalization techniques, Berkovits argued
that the sigma model still has an infinite number of conserved charges 
when quantum effects are taken into account \cite{Berkovits:2004xu}. 

Besides these general results, not much is known about the sigma model. 
The one loop conformal invariance was proved in \cite{BBHZZ,Brenno} and the 
argument for all loop conformal invariance was presented in \cite{Berkovits:2004xu}.
The one loop effective action was computed recently in \cite{Mazzucato:2009fv} where it was shown that the ``level'' of the CFT is not renormalized at one loop, which in turn means that the relation between the 't Hooft coupling $\lambda$ and the $AdS$ radius does not change at one loop. Besides, it was also shown that, using the prescription given in \cite{Berkovits:2004xu},  the effective action does not get any correction at all (neither local or non-local). Regarding the integrability of the model, a detailed study of the transfer matrix of the worldsheet was done in \cite{Mikhailov:2007mr}, where it was shown to be a well defined operator in quantum theory.  

In this work we consider the one-loop correction of the OPE's of the
left invariant currents.  This is one particularly interesting set of
operators in the worldsheet. Since these currents are not gauge
invariant they are not expected to be primary fields of the CFT,
nevertheless they are invariant under global $PSU(2,2|4)$
transformations and are used to construct integrated massless vertex
operators \cite{Mikhailovvertex} and also appear in massive unintegrated vertex operators.
Another complication is that these currents are not holomorphic even
in the classical limit, so their OPE's cannot be deduced from general
arguments. Therefore, if one wants to compute spacetime observables
using worldsheet techniques, a perturbative knowledge of these OPE's
is mandatory. Besides this practical application, the knowledge of
this current algebra in the worldsheet may shed light into more
general aspects of the theory, such as the apparent quantum
integrability. The tree level OPE's of these currents were computed in
\cite{Puletti:2006vb} (see also \cite{Mikhailov:2007mr} and
\cite{BianchiKluson}) while the algebra of the left and right currents for
a principal chiral model have been computed in \cite{ABT} and \cite{BenichouTroost}.

Surprisingly, most of the possible one-loop corrections vanish due to spacetime supersymmetry and the result obtained here corroborates with the effective action result obtained in \cite{Mazzucato:2009fv}. Thus 
this serves as further evidence that the relation between the 't Hooft coupling and the AdS radius is not renormalized. 

We also compute the OPE's of the left-invariant currents with the worldsheet 
energy momentum tensor. Although the currents are not primary fields, their 
tree level OPE with the energy momentum gives the expected result coming 
from gauge covariance. The results we found are compatible with general assumptions of CFT but they are not as simple as in the case of a chiral current algebra. Furthermore, at 1-loop we show that there is no 
correction to the tree level OPE for the bosonic currents. This is a 
surprising result since the left-invariant currents are not protected by any 
symmetry argument. On the other hand, the fermionic currents get anomalous dimension
contributions. However, this is not inconsistent, the two types of fermionic currents 
get contributions that cancel when combined into a single operator, so the stress 
energy tensor still has zero anomalous dimension. 


\paragraph{Organization} The structure of this paper is as follows. In
section two we review the
pure spinor superstring formalism. The case of $AdS_5 \times S^5$ background is discussed in section 3. In section four the methods to compute the OPE's is described. In section
five we compute the one-loop contributions to the OPE'S. Section 6 contains 
the computation of the OPE's between the left-invariant currents with the 
energy momentum tensor. In the section 7 we summarize and comment our
results. The appendices contain 
some technical details which were omitted in the main text. 

\section{Pure Spinor Type II Superstring in Curved Backgrounds}

In a curved background, the pure spinor sigma model action for the
type II superstring is obtained by adding to the flat action the integrated 
vertex operator for supergravity
massless states and then covariantizing with respect to ten dimensional
$N = 2$ super-reparameterization invariance. The result of this
procedure is

\be\label{action} S = {1\over{2\pi\a'}} \int d^2 z ~ ({1\over 2} \p Z^M \pb Z^N
(G_{NM} + B_{NM}) + d_\a  \pb Z^M E_{M}^\a  + \dt_{\ah}
\p Z^M E_{M}^{\ah} + \l^\a \o_\b \pb Z^M \O_{M\a}{}^\b  
\ee $$ + \lh^{\ah}
\oh_{\bh}\p Z^M \O_{M\ah}{}^{\bh} + d_\a \dt_{\bh} P^{\a\bh} + \l^\a \o_\b
\dt_{\gh} C_\a^{\b\gh} + \lh^{\ah} \oh_{\bh} d_\g
\Ct_{\ah}^{\bh\g} + \l^\a \o_\b  \lh^{\ah} \oh_{\bh}
S_{\a\ah}^{\b\bh} )+ S_{pure} + S_{FT},$$
where $E_M{}^A$ is the supervielbein
and $Z^M$ are the curved superspace coordinates, $B_{NM}$ is the
super two-form potential. $S_{pure}$ is the
action for the pure spinor ghosts and is the same as in the flat
space case. The pure spinor condition means that they
satisfy $\l^\a \g^{c} _{\a\b} \l^\b =0$ and $\lh^{\ah} \g^{c} _{\ah\bh}
\lh^{\bh} =0$, where $c = 0,{\ldots} 9$ is a tangent space bosonic index.

As was shown in \cite{BerkovitsHowe}, the gravitini and the dilatini
fields  are described by the lowest $\t$-components of the
superfields $C_{\a}^{\b\gh}$ and $\Ct_{\ah}^{\bh\g}$, while the
Ramond-Ramond field strengths are in the superfield $P^{\a\bh}$. The
dilaton is the theta independent part of the superfield $\Phi$ which
defines the Fradkin-Tseytlin term

\be\label{ft}
S_{FT} = {1\over{2\pi}} \int d^2z ~ r ~ \Phi,
\ee
where $r$ is
the world-sheet curvature. Because of the pure spinor constraints, the
superfields in (\ref{action}) cannot be arbitrary. In fact, it is
necessary that

\be\label{fields}\O_{M\a}{}^\b = \O_M ^{(s)} \d_\a{}^\b + {1\over 4} \O_{Mab}
(\g^{ab})_\a{}^\b,\quad \Ot_{M\ah}{}^{\bh} = \Ot_M ^{(s)} \d_{\ah}{}^{\bh}
+ {1\over 4} \Ot_{Mab} (\g^{ab})_{\ah}{}^{\bh}, \ee
$$
C_\a^{\b\gh} = C^{\gh} \d_\a{}^\b + {1\over 4} C_{ab}{}^{\gh}
(\g^{ab})_\a{}^\b,\quad \Ct_{\ah}^{\bh\g} = \Ct^\g
\d_{\ah}{}^{\bh} + {1\over 4} \Ct_{ab}{}^\g
(\g^{ab})_{\ah}{}^{\bh},$$
$$
S_{\a\ah}^{\b\bh} = S \d_\a{}^\b \d_{\ah}{}^{\bh} + {1\over 4}
S_{ab} (\g^{ab})_\a{}^\b \d_{\ah}{}^{\bh} + {1\over 4} \St_{ab}
(\g^{ab})_{\ah}{}^{\bh} \d_\a{}^\b + {1\over{16}} S_{abcd}
(\g^{ab})_\a{}^\b (\g^{cd})_{\ah}{}^{\bh}.$$

The engineering dimensions, i.e dimensions in units of space-time
length, for the worldsheet fields in (\ref{action}) are: 
\be
\label{wsdim}
[X^m]  =1, \,\, [\theta^\mu ]= {1\over 2}, \,\, [d_\a] = [\dt_{\ah}] =
{3\over 2},
\,\, [\l^\a
\o_\b ] =  [ \lh^{\ah}  \oh_{\bh}]= 2 .
\ee

\section{Review of the Pure Spinor Superstring in $AdS_5 \times S^5$}
As was shown for the first time in \cite{MetsaevTseytlin}, the
superstring in $AdS_5 \times S^5$ background can be described using
some currents defined in the superalgebra $psu(2,2|4)$. Those
currents, which are defined in a left-invariant way, are given by $J^A
= (g^{-1} \p g)^A = \p Z^M E_{M}^A $, $\Jb^A
= (g^{-1} \pb g)^A = \pb Z^M E_{M}^A$ for $g$ an element in the coset supergroup
$PSU(2,2|4)/SO(4,1)\times SO(5)$. The index $A$ denotes $(\au, \a, \ah,
)$ and $a = 0,{\ldots} 4$ for $AdS_5$, $a' = 5,{\ldots} 9$ for $S^5$, $\a = 1,{\ldots}
16$, $\ah = 1,{\ldots} 16$ and $\au$ denotes both $a$ and $a'$. 

Another way of obtaining the action for the superstring in the $AdS_5
\times S^5$ background is by replacing the values that the superfields of
the action (\ref{action}) take on that background, as shown in
\cite{BerkovitsCQS} and \cite{BerkovitsChandia}. In the following we
will review that procedure. 

Using the supervielbein and the definition of the currents given
above, one can check that the term which contains $G_{MN}$ can be
written as 
\be\label{G}{1\over 2} \p Z^M \pb Z^N G_{NM} = {1\over 2} J^{\au} \Jb^{\bu} \eta_{\au\bu}.
\ee
In $AdS_5\times S^5$ the only non-zero component of $B_{NM}$ is
$B_{\a\bh} = {1\over 2} (N g_s)^{1\over 4} \sqrt{\a'} \d_{\a\bh}$, where 
$\d_{\a \bh} = (\g^{01234})^{\a\bh}$. Then the
term containing $B_{NM}$ in the action will lead to 
\be\label{B}{1\over 2} \p Z^M \pb Z^N B_{NM} = {1\over 2} (J^{\bh} \Jb^\a B_{\a\bh} +
J^\a \Jb^{\bh} B_{\bh \a}) ={1\over 4} \sqrt{\a'} (N g_s)^{1\over 4} 
(J^{\bh} \Jb^\a + J^\a \Jb^{\bh})\d_{\a\bh}.
\ee
From the definitions of the currents $J^A$, the terms containing
explicitly $E_M ^\a$ and $E_M ^{\ah}$  in (\ref{action}) will give
\be\label{E}d_\a \pb Z^M E_M ^\a = d_\a \Jb^\a , \,\, \dt_{\ah} \p Z^M
E_M ^{\ah} = \dt_{\ah} J^{\ah}.
\ee
By computing the flux of the five-form Ramond-Ramond field-strength
one finds that
\be\label{RRflux}P^{\a\bh} = {\d^{\a\bh} \over {\sqrt{\a'}(N g_s)^{1\over
4}}},
\ee
where $\d^{\a\bh} = (\g^{01234})^{\a\bh}$ and actually (\ref{RRflux}) sets the value for $B_{\a\bh}$ written above, as can be 
proven by
using the field-strenth $H = dB$ and the constraints of
\cite{BerkovitsHowe} . The values of the Superfields $C_\a ^{\b\gh}$ and
$\Ct_{\ah}^{\bh\g}$ are zero in the $AdS_5 \times S^5$, as well as
$\O_M ^{(s)}$ and $\Ot _M ^{(s)}$ because they are related to
derivatives of the superfield containing the dilaton, which is
constant for this background. 
Now, the terms containing the spin connections will lead to 

\be\label{spinconnections}\l^\a \o_\b \pb Z^M \O_{M\a}{}^\b = 
N_{\au\bu} \Jb^{\au\bu} , \,\, \lh^{\ah} \oh_{\bh} \p Z^M
\Ot_{M\ah}{}^{\bh} = \hat N_{\au\bu} J^{\au\bu} , 
\ee
where $J^{\au\bu} = {1\over 2} \p Z^M \Ot_{M\au\bu}$, $\Jb^{\au\bu} = {1\over 2}
\pb Z^M \O_{M\au\bu}$ and $N^{\au \bu} = {1\over 2}
(\l \g^{\au \bu} \o)$,  $\Nh^{\au \bu} = {1\over 2}
(\lh \g^{\au \bu} \oh)$ are the pure spinors Lorentz currents.
Finally, the term containing $S_{\a\ah}^{\b\bh}$ is related to the
space-time curvature as shown in \cite{BerkovitsHowe}, which is constant
for the $AdS_5 \times S^5$ space. More specifically, 

\be\label{Scurvature}R_{abcd} = -{1\over R^2} \eta_{a[c}\eta_{d]b}   \,\,\, 
R_{a'b'c'd'} = {1\over R^2} \eta_{a'[c'}\eta_{d']b'},
\ee
where $R$  is the radius of $AdS_5$ and $S^5$ and the minus sign in
the first equation is because $AdS_5$ has negative curvature. Thus,
replacing the values of the background fields given above, the
following action is found

\be\label{actioninmzo}S = {1\over {2\pi\a' }} \int d^2 z ({1\over 2 } J^{\au} \Jb ^{\bu}
\eta_{\au \bu} + {1\over 4}(Ng_s)^{1\over 4}\sqrt{\a'}\d_{\a\bh} (J^\a
\Jb^{\bh} +  J^{\bh} \Jb ^\a) +d_\a \Jb^\a + \dt_{\ah} J^{\ah} 
\ee
$$+
{1\over (Ng_s)^{1\over 4} \sqrt{\a'}}d_\a \dt_{\bh} \d^{\a\bh} + 
N_{\au \bu}
\Jb^{[\au \bu]} + \Nh_{\au \bu}J^{[\au \bu]} - {1\over R^2}N_{ab}
\Nh^{ab} + {1\over R^2}N_{a'b'} \Nh^{a'b'} ) +
S_\l + S_{\lh}.$$
Note by now that the engineering dimensions are
\be\label{Jdim}[J^{\au}] = [\Jb^{\au}] = 1, \,\, [J^\a] = [J^{\ah}] =
[\Jb^\a] = [\Jb^{\ah}] = {1\over 2} ,
\ee
$$ [N_{\au\bu}] = [\hat N_{\au\bu}]
= 2 , \,\, [d_\a] = [\dt_{\ah}] = {3\over 2}, [J^{\au\bu}] = [\Jb^{\au\bu}] = 0.
$$
By defining $\a^{-1} = (Ng_s)^{1\over 4}$, using the equations of
motion for $d_\a$ and $\dt_{\ah}$ and performing the scalings
\be\label{scalings}
(J^{\au} , \Jb^{\au}) \to \a^{-1}(J^{\au} ,\Jb^{\au}),
\,\,\, (J^{\a}, J^{\ah} , \Jb^\a ,\Jb^{\ah}) \to 2 {(\a')^{1\over 4}
\over \sqrt{\a}}(J^{\a} ,J^{\ah} , \Jb^\a ,\Jb^{\ah})
\ee
$$ 
(N_{\au\bu} , \hat N_{\au\bu}) \to {1\over{\a R} }(N_{\au\bu} , \hat
N_{\au\bu} ) , \,\,\, (J^{\au\bu} , \Jb^{\au\bu}) \to {\sqrt{\a'}\over
\a} (J^{\au\bu} , \Jb^{\au\bu}) ,
$$
we find the action
\be\label{adsaction}S = {1\over {2\pi\a'\a^2}} \int d^2 z ({1\over 2 } J^{\au} \Jb ^{\bu}
\eta_{\au \bu} + \d_{\a\bh} (J^\a \Jb^{\bh} - 3 J^{\bh} \Jb ^\a) \ee $$ +
N_{\au \bu}
\Jb^{[\au \bu]} + \Nh_{\au \bu}J^{[\au \bu]} - N_{ab} \Nh^{ab} +
N_{a'b'} \Nh^{a'b'} ) +
S_\l + S_{\lh},
$$
which coincides with ``usual'' action for the superstring written in
terms of the $psu(2,2|4)$ currents \cite{BerkovitsCQS}
\cite{BerkovitsChandia}. Note also that in (\ref{adsaction}) all
$J$'s, $\Jb$'s, and pure spinor Lorentz currents has engeneering
dimension one. So, by choosing units in which $2\pi \a' =1$ the action
is given in terms of dimensionless worldsheet fields.

Because of their definition, $(J^A , \Jb^A)$ satisfy
the Maurer-Cartan identities $\p \Jb^A - \pb J^A + [J,\Jb]^A = 0$, so
by making a variation of the action and using those identities, we can
find the equations of motion 
\be
\label{eom}\N \Jb_2 = -[J_1 , \Jb_1] + {1\over 2} [N,\Jb_2] - {1\over
2}[J_2 , \hat N], 
\ee
\be \label{eomJ2}
\bar\N J_2 = [J_3 , \Jb_3] - {1\over 2} [J_2,\hat N] + {1\over
2}[N , \Jb_2] 
\ee
\be
\N \Jb_1 = {1\over 2} [N,\Jb_1]  - {1\over 2} [J_1 , \hat N],
\ee
\be
\Nb J_1 = [J_2 , \Jb_3] + [J_3, \Jb_2] + {1\over 2} [N, \Jb_1] -
{1\over 2} [J_1 , \hat N]
\ee
\be
\Nb J_3 = {1\over 2} [N,\Jb_3]  - {1\over 2} [J_3 , \hat N],
\ee
\be
\N \Jb_3 = -[J_2 , \Jb_1] - [J_1, \Jb_2] + {1\over 2} [N, \Jb_3] -
{1\over 2} [J_3 , \hat N],
\ee
where $\N = \p + [J_0 ,\, ]$ and $\bar\N = \pb + [\Jb_0 ,\, ]$. We have 
supressed the index $A$ and introduced a sub-index
$0,1,2,3$ for the currents. This notation stands for $J_0 = J^{[\au \bu]}
M_{\au \bu}$, $J_1 = J^\a Q_\a$, $J_2 = J^{\au} P_{\au}$, $J_3 = J^{\ah} {\hat
Q}_{\ah}$
and similarly for the $\Jb$ currents. This $Z_4$ gradding for the
superalgebra was noted in \cite{BBHZZ}. Note that we have written the currents in terms of the generators of
$psu(2,2|4)$, whose structure constants different from zero are

\be\label{structureconstants}f^{\cu} _{\a\b} = 2 \g^{\cu} _{\a\b} , \,\,\,\, f^{\cu}
_{\ah\bh} = 2 \g^{\cu} _{\ah\bh} \ee
$$
f^{[ef]}_{\a\bh} = (\g^{ef})_\a {}^\g \d_{\g\bh} = - (\g^{ef})_{\bh}
{}^{\gh} \d_{\a\gh} = f^{[ef]}_{\bh\a},
\,\,\, f^{[e'f']}_{\a\bh} = -(\g^{e'f'})_\a {}^\g
\d_{\g\bh}  = (\g^{e'f'})_{\bh} {}^{\gh}
\d_{\a\gh}  = f^{[e'f']}_{\bh\a}, $$ $$
f^{\bh} _{\a \cu} = - f^{\bh} _{\cu\a} = {1\over 2} (\g_{\cu})_{\a\b} \d^{\b\bh} ,
\,\,\, f^\b _{\ah \cu} = - f^\b _{\cu \ah} = -{1\over 2} (\g_{\cu})_{\ah\bh}
\d^{\b\bh},$$ $$
f_{cd}^{[ef]} = {1\over 2} \d_c ^{[e} \d_d ^{f]} , \,\,\,
f_{c'd'}^{[e'f']} = -{1\over 2} \d_{c'} ^{[e'} \d_{d'} ^{f']}, $$ $$
f_{[\cu \du][\eu \fu]}^{[\gu \hu]} = {1\over 2} (\eta_{\cu \eu} \d_{\du} ^{[\gu} \d_{\fu} ^{\hu]} -
\eta_{\cu \fu} \d_{\du} ^{[\gu} \d_{\eu} ^{\hu]}+\eta_{\du \fu} \d_{\cu} ^{[\gu} \d_{\eu} ^{\hu]} -
\eta_{\du \eu} \d_{\cu} ^{[\gu} \d_{\fu} ^{\hu]}),$$ $$
f^{\fu} _{[\cu \du] \eu} = - f^{\fu} _{\eu [\cu \du]} = \eta_{\eu[\cu}\d_{\du]}^{\fu} , \,\, f^\b
_{[\cu \du]\a} = -f^\b _{\a [\cu \du]} = {1\over 2} (\g_{\cu \du})_\a {}^\b , \,\,
f^{\bh}
_{[\cu \du]\ah} = -f^{\bh} _{\ah [\cu \du]} = {1\over 2} (\g_{\cu \du})_{\ah} {}^{\bh} .
$$
The pure spinors have
also equations of motion, given by $\Nb N = {1\over 2} [N,\hat N]$ and $\N
\hat N = {1\over 2} [N,\hat N]$.

\section{OPE'S in momentum space and dimensional regularization}

In this section it is described the kind of calculations we intend to do. We are going to calculate contributions to the expectation values
 $\left\langle J^{\au}(y)J^{\bu}(z)\right\rangle$, $\left\langle
 J^{\au}(y)J^{\a}(z)\right\rangle$, etc... perturbatively, including double contractions (one  loop) with no contributions of classical fields. The traditional way to calculate this kind of expectation values is to perform 
 a background field expansion as in \cite{BBHZZ}, 
\cite{Brenno} and \cite{BedoyaYM}.
That is, we choose a classical background
given by an element $g_0$ in the supergroup and parametrize the
quantum fluctuations by $X$ as $g = g_0 e^{\a X}$, where $\a$ is the
coupling constant defined in the last section. Then, the currents
can be written as 
\be\label{currentsplit}
J = g^{-1}\p g = e^{-\a X}J_0 e^{\a X} +  e^{-\a
    X}\p e^{\a X}, \ee
    $$ 
\Jb = g^{-1}\pb g = e^{-\a X}\Jb_0 e^{\a X} +  e^{-\a
    X}\pb e^{\a X}. $$
The exponentials in (\ref{currentsplit}) can be expanded, giving rise to
expressions involving commutators, which can be evaluated by using
the structure constants of the $psu(2,2|4)$ Lie superalgebra
(\ref{structureconstants}), that is, 

\be\label{Jsplit}J = J_0 + \a (\p X + [J_0 , X]) + {{\a^2}\over 2}([\p X ,
  X] + [[J_0,X],X]) + {{\a^3}\over {3!}}([[\p X , X],X]+[[[J_0
  ,X],X],X]) + ... ,
  \ee
and similarly for $\Jb$.\footnote{ Note that we have made the choice $X= X_2 + X_1 +X_3$ for the parametrization of the coset. Here we have used the $SO(1,4)\times SO(5)$ gauge invariance to fix $X_0=0$. Supposed we do not use the gauge invariance to fix this component and use another parametrization $X'= X'_2+X'_1+X'_3+X'_0$. We can use the Baker-Campbell-Hausdorff formula to write $e^{X'} = e^{X_2 + X_1 +X_3 } e^{X_0}$ and find the field redefinitions from from $X'$ to $X_2 + X_1 +X_3$ and $X_0$. If we define the quantum field by $g=g_0 e^{X_2 + X_1 +X_3 } e^{X_0}$ the expanded action will be independent of $X_0$, so this component is just a local gauge transformation of currents $J_i \to e^{-X_0} J_ie^{X_0}$ for $i =1,2$ and $3$.  This implies that a coset parametrization without fixing $X_0$ to vanish is related to our choice by a gauge transformation, as it should. Although our results are not gauge invariant, they are gauge covariant, so we do not expect any significant change using another coset parametrization.} In the last expression $J_0$ denotes the
classical part of $J$ and not the index of the $Z_4$ gradding. That
sub-index will be dropped out, so it will be understood that the
currents
which appears in this type of expansion are classical. In the
appendix, the expansion of the terms in the action (\ref{adsaction}) is
written up to cubic terms in the quantum fields, since this is the
relevant order  for the one-loop computation of the current's OPE's.  We will focus on the matter part of the OPE's. In Section 5 they do not enter at all, since there is no diagram that mixes matter with ghosts. However, in Section 6, they do enter at tree and one loop level. 
 
Replacing those expansions of the appendix in (\ref{adsaction}), one can identify the kinetic 
piece $S_p$ of the action 
\be\label{Sp}S_p = \int d^2 z ({1\over 2} \p X^{\au} \pb X^{\bu}
  \eta_{\au\bu} + 4 \d_{\a\bh}\p X^\a \pb X^{\bh} ),
  \ee
from which we obtain the propagators in coordinate space 

\begin{equation}\label{propb}
X^a (y) X^b (z) \rightarrow -\eta^{ab} \ln |y-z|^2 , \:\:\:\:
X^{a'} (y) X^{b'} (z) \rightarrow -\d^{a'b'} \ln |y-z|^2 
\end{equation}
$$
X^{\alpha} (y) X^{\hat\beta} (z) \rightarrow -\frac{1}{4}
\delta^{\alpha\hat\beta} \ln |y-z|^2 .
$$

The reminder terms of the background expansion will provide the
vertices of the theory. It is then straightforward to write down
coordinate-space   expressions for the Feynman rules of the diagrams
which will appear in the remaining of the paper, and calculate the 
contribution of each OPE, like the tree level calculations of \cite{Puletti:2006vb}. However, things are different at one loop. There are divergences which 
produce ambiguities in the coodinate-space integrals. The basic techniques 
for dealing with such a problem, involving this kind of calculation,  were 
developed a long time ago in \cite {boer}, \cite {skenderis}, \cite{nedel}, 
when it were used  momentum space Feynman rules with a prescription for 
worldsheet dimensional regularization. Then the results could be
written in coordinate space by using an inverse Fourier transformation.

The two dimensional prescription for dimensional regularization consists in keeping all the interactions in exactly two dimensions, but the kinetic terms, and hence the denominators of the propagators will be in d=$2-2\epsilon$ dimensions. 
  
  We are going to use the definition $d^2 k= \frac{dk_xdk_y}{\pi}$. With this choice there is no $\pi$ dependence in the results and the Green function $G(y,z)$ is represented as
  \begin{equation}
  G(y,z)= \int d^2k\frac{e^{ik(y-z)+i\bar{k}(\bar{y}-\bar{z})}}{\left.k\right.^2}.
  \end{equation}
              
The  momentum space propagators  look like

\be\label{propms}X^{\au} (k) X^{\bu} (l) \rightarrow \eta^{\au \bu}{\d^2
(k+l)\over{|k|^2}}, \,\,\, X^\a (k) X^{\bh} (l) \rightarrow {1\over
4}\d^{\a\bh}{\d^2 (k+l)\over{|k|^2}}.
\ee

To work out the corresponding expression for the OPE´s in  momentum space we use the dimensional regularization prescription and include a factor $\Gamma(1-\epsilon)(4\pi)^{-\epsilon}(2\pi)^{2\epsilon}$ for each loop. 
This will remove the Euler constant (the G-scheme \cite {gesquem}).
All the integrals we need to compute in the momentum space come from
the formula
\begin{eqnarray}
&&\int d^{d}p\frac{%
p^{a}\overline{p}^{b}}{[\left| p\right| ^{2}]^{\alpha }[\left|
p-k\right|^2
]^{\beta }} = \nonumber \\
&&k^{a+1-\alpha -\beta }\overline{k}^{b+1-\alpha -\beta }\left| \frac{k^{2}}{%
\mu ^{2}}\right| ^{-\epsilon }
 \times \sum_{i=0}^{i=a}\left(
\begin{array}{c}
a \\
i
\end{array}
\right) [\frac{\Gamma \left( 2-\alpha -\beta +b+i-\epsilon \right) }{%
\Gamma \left( 2-2\epsilon -\alpha -\beta +i+b\right) }\nonumber \\
&&\times \frac{\Gamma \left( \alpha +\beta -1-i+\epsilon \right)}{\Gamma \left( 1+\epsilon \right) \mu ^{-2\epsilon }} \Gamma \left(
1-\epsilon -\beta +i\right) ] \label{regul},
\end{eqnarray}
where $\mu$ is the usual mass parameter of the dimensional
regularization and the measure $d^dp$ is the standart d-dimensional measure divided by $\pi$.  Using this regularization, integrals like $\int \frac{d^dk}{|k|^2}$
vanish due to the cancelation betwen ultraviolet and infrared divergences in two dimensions. In order to check whether infrared and ultraviolet divergences cancel separately, we should replace the propagator in each infrared 
diagram by \cite{colina, grisaru, boer}
\begin{equation}
\frac{1}{\left|k\right|^2} \rightarrow  \frac{1}{\left|k\right|^2} + \frac{1}{\eta}\d^ 2(k)
\end{equation}
and by taking $\epsilon = \eta$ we could subtract out all infrared divergences. Since we are not evaluating expectation values of conserved currents, the result may depend on $\epsilon$ and  this procedure is important. However, for the sake of simplicity  we are not going to do this in this paper and we  postpone this discussion to a future work.\footnote{If we keep the parameters $\epsilon$ and  $\eta$ as independent parameters the infrared divergences can be read from the  $\frac{1}{\epsilon} - \frac{1}{\eta}$ coefficients and the ultravilolet from $\frac{1}{\epsilon}$ coefficients.}

Next,  we need to calculate all diagrams in momentum space using (\ref{regul}) with the dimensional regularization prescription,  and afterwards reexpress the results in coordinate space using the following:
\begin{equation}\label{fourier}
\frac{k}{\overline{k}} \longleftrightarrow -\frac{1}{\left( y-z\right) ^{2}%
},\qquad \frac{\overline{k}}{k}\longleftrightarrow -\frac{1}{\left( \overline{%
y}-\overline{z}\right) ^{2}},
\end{equation}
$$\frac{k}{\varepsilon \overline{k}}+\frac{k}{\overline{k}}\left( 1-\log \frac{%
\left| k\right| ^{2}}{\mu ^{2}}\right) \longleftrightarrow -\frac{\ln
|y-z|^2}{\left( y-z\right) ^{2}},  $$
$$\frac{\overline{k}}{\varepsilon k}+\frac{\overline{k}}{k}\left( 1-\log \frac{%
\left| k\right| ^{2}}{\mu ^{2}}\right) \longleftrightarrow -\frac{\ln |y-z|^2}{\left( \overline{y}-\overline{z}\right) ^{2}}.  $$

\section{OPE'S without classical part }

First note that from the expansions in the appendix, collecting the terms with three quantum
fields and no classical field we obtain

\be\label{SXXX}S(X^3) = {\a \over 4} \int d^2 z[ \p X^{\au} \pb X^\a X^\b
(\g_{\au})_{\a\b} -  \p X^{\au} \pb X^{\ah} X^{\bh}
(\g_{\au})_{\ah\bh} -  \pb X^{\au} \p X^\a X^\b
(\g_{\au})_{\a\b} +  \pb X^{\au} \p X^{\ah} X^{\bh}
(\g_{\au})_{\ah\bh} \ee
$$  +2 X^{\au} \p X^\a \pb X^\b (\g_{\au})_{\a\b}
- 2 X^{\au} \p X^{\ah} \pb X^{\bh} (\g_{\au})_{\ah\bh} ]. $$
Integrating by parts the first line we obtain
\be\label{SXXXa}S(X^3) = \a \int d^2 z  [ X^{\au} \p X^\a \pb X^\b
(\g_{\au})_{\a\b} -   X^{\au} \p X^{\ah} \pb X^{\bh}
(\g_{\au})_{\ah\bh} ].
\ee 
The last expression gives the vertices used in the computation
detailed in the next subsection. 
\subsection{One-loop computations}
We can use the expansions of the appendix to compute perturbatively
the OPE's of the various currents $J^{A}$ and  $\Jb^A$. We will give
in detail the computation of $\langle J^{\au} (y) J^{\bu}(z)\rangle $ leaving the method
clear and explaining how to get the rest of the results.

Restricting the expansion (\ref{Jsplit}) to the case without classical
currents, we can write

\be\label{J_0J_0} \langle J^{\au} (y) J^{\bu}(z)\rangle = \a^2 \langle
\p X^{\au}(y) \p X^{\bu} (z) \rangle - \a^3 \langle \p X^{\au} (y) \p
X^\a  X^{\b} (z)\rangle \g^{\bu}_{\a\b} - \a^3 \langle \p X^{\au} (y) \p
X^{\ah}  X^{\bh} (z)\rangle \g^{\bu}_{\ah\bh}
\ee 
$$ - \a^3 \langle \p X^\a  X^{\b} (y) \p X^{\bu}(z)\rangle \g^{\au}_{\a\b}
- \a^3 \langle \p X^{\ah}  X^{\bh} (y) \p X^{\bu}(z)\rangle
\g^{\au}_{\ah\bh} + \a^4 \langle \p X^\a X^\b (y) \p X^{\gh} X^{\dh} (z)
\rangle \g^{\au}_{\a\b} \g^{\bu}_{\gh\dh} $$ $$+ \a^4 \langle \p X^{\ah}
X^{\bh} (y) \p X^{\g} X^{\d} (z)
\rangle \g^{\au}_{\ah\bh} \g^{\bu}_{\g\d} .$$
With the first term in (\ref{J_0J_0}) we can form a one-loop diagram
by using the two terms in the right hand side of (\ref{SXXXa}), which
will come from the expansion of the exponential of minus the action at
second order. This one-loop diagram is shown below \footnote{In all the diagrams crosses indicate vertices coming from the currents, double lines indicate background fields and single lines indicate quantum fields.} .

\be \epsfbox{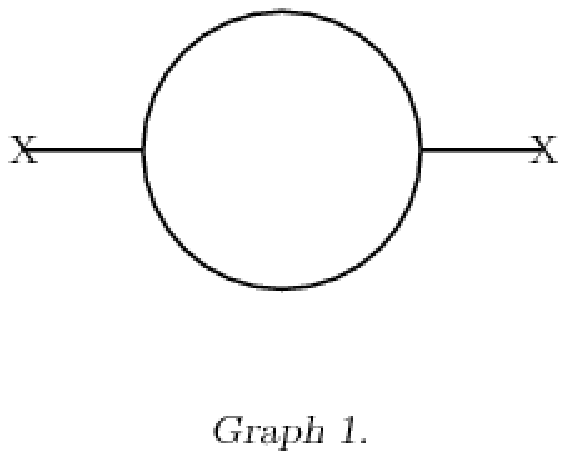} \label{picII} \ee
So, in momentum space, using the contractions (\ref{propms}) the first
diagram gives

\be \a^2 \langle \p X^{\au}(y) \p X^{\bu} (z) \rangle = -{\a^4 \over
8} \eta^{\au \cu}\eta^{\bu\du} (\g_{\cu})_{\a\b}(\g_{\du})_{\ah\bh}
[\d^{\a\bh}\d^{\b\ah} \int d^d p {p\bar p \over |p|^2} {(k-p)(\bar k
-\bar p) \over |k-p|^2 } 
\ee $$ - \d^{\a\ah}\d^{\b\bh}\int d^d p {p^2 \over |p|^2} {(\bar k -
\bar p)^2 \over |k-p|^2}] \left ({k\over |k|^2} \right)^2 .
$$
The coefficient deserves an explanation. There is a ${1\over 2}$
coming from the expansion of $exp{-S}$ at the second order in $S$,
also there is symmetry factor of $2$ from the different possibilities
of contracting the bosonic indices. Another factor of two comes from
the double product in (\ref{SXXXa}) when computing $S(X^3)^2$.
Finally, there is a ${1\over 4^2}$ from the fermionic propagator.
It can be easily checked that $(\g_{\cu})_{\a\b}(\g_{\du})_{\ah\bh}
\d^{\a\ah}\d^{\b\bh} = 16\eta_{\cu\du}$. Therefore, using the results
of the integrals summarized in the appendix, we obtain

\be \label{J_0J_0I}\a^2 \langle \p X^{\au}(y) \p X^{\bu} (z) \rangle = -2 \a^4
\eta^{\au\bu}[{k\over \bar k} + {k\over \bar k}({1\over \epsilon} + 1 -
ln{|k|^2 \over \mu^2})].
\ee
Now let's consider the remaining terms in (\ref{J_0J_0}). Both the second
and third terms in (\ref{J_0J_0}) can be represented by the diagram
\be \epsfbox{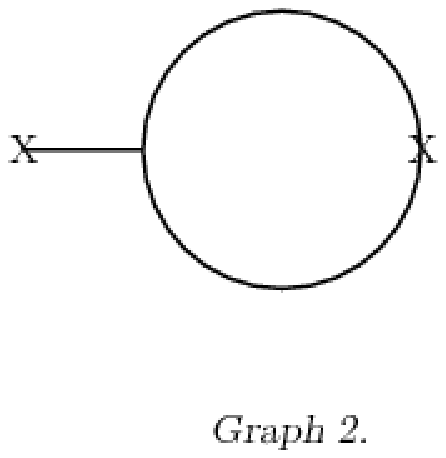} \label{picII} \ee
It can be checked that this diagram cancels because the second term 
in (\ref{J_0J_0}) cancels with the third. The reason for this
cancelation is as follows: to form the diagram the second term in
(\ref{J_0J_0}) contracts with the second term
in (\ref{SXXXa}), while the third term in (\ref{J_0J_0}) contracts
with the first term in (\ref{SXXXa}). Since those terms in (\ref{SXXXa})
have opposite signs then $-\a^3 \langle \p X^{\au} (y) \p
X^\a  X^{\b} (z)\rangle \g^{\bu}_{\a\b}$ cancels with $ - \a^3 \langle \p X^{\au} (y) \p
X^{\ah}  X^{\bh} (z)\rangle \g^{\bu}_{\ah\bh}$. Using the same
reasoning one can check that $- \a^3 \langle \p X^\a  X^{\b} (y) \p X^{\bu}(z)\rangle \g^{\au}_{\a\b}
$ cancels with $- \a^3 \langle \p X^{\ah}  X^{\bh} (y) \p X^{\bu}(z)\rangle
\g^{\au}_{\ah\bh}$. That means the following diagram also cancels
\be \epsfbox{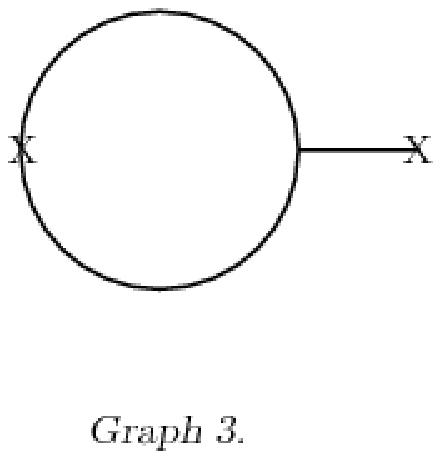} \label{picIII}\ee

Finally the term $\a^4 \langle \p X^\a X^\b (y) \p X^{\gh} X^{\dh} (z)
\rangle \g^{\au}_{\a\b} \g^{\bu}_{\gh\dh}$ in (\ref{J_0J_0}) , which
is represented by the diagram

\be \epsfbox{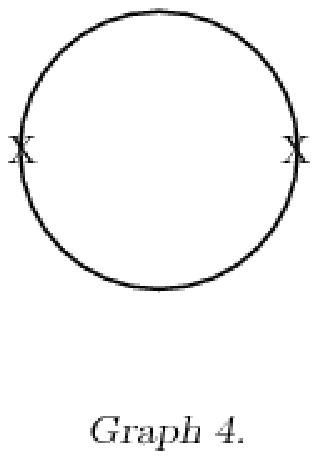} \label{picIV}  \ee
gives

\begin{eqnarray}\nonumber
\a^4 \langle \p X^\a X^\b (y) \p X^{\gh} X^{\dh} (z)
\rangle \g^{\au}_{\a\b} \g^{\bu}_{\gh\dh} &=& {\a^4 \over 4^2}
\g^{\au}_{\a\b}\g^{\bu}_{\gh\dh} [- \d^{\a\gh}\d^{\b\dh} \int d^d p{p^2
\over |p|^2} {1\over |k-p|^2} \\ \nonumber& +& \d^{\a\dh}\d^{\b\gh}\int d^d p {p\over
|p|^2} {(k-p)\over |k-p|^2}] \\ \label{J_0J_0IV}
&=& \a^4 \eta^{\au\bu} \left[{k\over \bar k} + {k\over \bar k}\left({1\over \epsilon} + 1 -
ln{|k|^2 \over \mu^2}\right)\right]
\end{eqnarray}
and the last term in (\ref{J_0J_0}) gives the same result. Because of
this fact, (\ref{J_0J_0I}) cancels with two times the result in
(\ref{J_0J_0IV}), or in other words, the first diagram cancels with
the forth. Then, the one-loop correction to $\langle J^{\au} (y)
J^{\bu} (z) \rangle $ without classical field vanishes.

Let's consider now $\langle J^{\a}(y) J^{\bh} (z) \rangle$ at one loop
and also without classical currents contributions. Then

\be\label{J_1J_3} \langle J^{\a} (y) J^{\bh}(z)\rangle = \a^2 \langle
\p X^{\a}(y) \p X^{\bh} (z) \rangle + {1\over 4}\a^3 \langle \p X^{\a} (y) \p
X^\g  X^{\bu} (z)\rangle (\g_{\bu})_{\g\d}\d^{\d\bh} 
\ee 
$$- {1\over 4}\a^3 \langle 
\p X^{\a} (y) \p 
X^{\bu}  X^{\g} (z)\rangle (\g_{\bu})_{\g\d}\d^{\d\bh}
 - {1\over 4}\a^3 \langle \p X^{\gh}  X^{\bu} (y) \p X^{\bh}(z)\rangle
 (\g_{\bu})_{\gh\dh}\d^{\a\dh} $$ $$+ {1\over 4}\a^3 \langle \p X^{\bu}
 X^{\gh} (y) \p X^{\bh}(z)\rangle
 (\g_{\bu})_{\gh\dh}\d^{\a\dh} 
 - {1\over 16} \a^4 \langle \p X^{\gh} X^{\cu} (y) \p X^{\g} X^{\du} (z)
\rangle (\g_{\cu})_{\gh\dh}\d^{\a\dh} (\g_{\du})_{\g\d}\d^{\d\bh} $$
$$+ {1\over 16} \a^4 \langle \p X^{\gh} X^{\cu} (y) \p X^{\du} X^{\g} (z)
\rangle (\g_{\cu})_{\gh\dh}\d^{\a\dh} (\g_{\du})_{\g\d}\d^{\d\bh}
+{1\over 16}
\a^4 \langle \p X^{\cu} X^{\gh} (y) \p X^{\g} X^{\du} (z)
\rangle (\g_{\cu})_{\gh\dh}\d^{\a\dh} (\g_{\du})_{\g\d}\d^{\d\bh} $$ $$
-{1\over 16} \a^4 \langle \p X^{\cu} X^{\gh} (y) \p X^{\du} X^{\g} (z)
\rangle (\g_{\cu})_{\gh\dh}\d^{\a\dh} (\g_{\du})_{\g\d}\d^{\d\bh} .$$

The result will be analog in this case. The first term, represented by
diagram 1 gives
\be \label{J_1J_3I}\a^2 \langle \p X^{\a}(y) \p X^{\bh} (z) \rangle =
{5\over 16} \a^4
\d^{\a\bh}[ {k\over {\bar k}} +{k\over \bar k}({1\over \epsilon} + 1 -
ln{|k|^2 \over \mu^2})],
\ee
while the second term cancels the third in (\ref{J_1J_3}), as well as
the fourth cancels the fifth. Again, the second and third diagrams
cancel independently. Also, in this case those cancellations
are due to the sign difference in the two terms of (\ref{SXXXa}).
The last four terms in (\ref{J_1J_3}) are represented by the fourth
diagram. The sixth term in (\ref{J_1J_3}) gives
\be\label{J_1J_3II}
 - {1\over 16} \a^4 \langle \p X^{\gh} X^{\cu} (y) \p X^{\g} X^{\du} (z)
\rangle (\g_{\cu})_{\gh\dh}\d^{\a\dh} (\g_{\du})_{\g\d}\d^{\d\bh} =
-{5\over 32}\a^4 \d^{\a\bh} {k\over \bar k}({1\over \epsilon} + 1 -
ln{|k|^2 \over \mu^2}),
\ee
which is also the result of the eighth term. Finally, the seventh term in
(\ref{J_1J_3}) gives
\be\label{J_1J_3III}
 {1\over 16} \a^4 \langle \p X^{\gh} X^{\cu} (y) \p X^{\du} X^{\g} (z)
\rangle (\g_{\cu})_{\gh\dh}\d^{\a\dh} (\g_{\du})_{\g\d}\d^{\d\bh} =
-{5\over 32}\a^4 \d^{\a\bh}{k\over \bar k},
\ee
which is the same result for the nineth term. Then, twice (\ref{J_1J_3II}) plus 
twice (\ref{J_1J_3III}) cancels with (\ref{J_1J_3I}), or again, the
first diagram cancels with the fourth.

Let's now consider $\langle J^{\au\bu} (y) J^{\cu\du} (z)\rangle$. For
this case only diagram 4 contributes.
\be
\langle J^{\au\bu} (y) J^{\cu\du} (z)\rangle = {\a^4 \over 4} \langle
\p X^\a X^{\bh} (y) \p X^\b X^{\ah}(z)\rangle (\g^{\au\bu})_\a {}^\g
\d_{\g\bh} (\g^{\cu\du})_\b {}^\d \d_{\d\ah} 
\ee
$$- {\a^4 \over 4} \langle
\p X^\a X^{\bh} (y) \p X^{\ah} X^{\b}(z)\rangle (\g^{\au\bu})_\a {}^\g
\d_{\g\bh} (\g^{\cu\du})_{\ah} {}^{\dh} \d_{\b\dh} - {\a^4 \over 4} \langle
\p X^{\bh} X^{\a} (y) \p X^{\b} X^{\ah}(z)\rangle (\g^{\au\bu})_{\bh}
{}^{\gh}\d_{\a\gh} (\g^{\cu\du})_{\b} {}^{\d} \d_{\d\ah}$$ $$
+{\a^4 \over 4} \langle
\p X^{\bh} X^{\a} (y) \p X^{\ah} X^{\b}(z)\rangle (\g^{\au\bu})_{\bh}
{}^{\gh}
\d_{\a\gh} (\g^{\cu\du})_{\ah} {}^{\dh} \d_{\b\dh} + {\a^4 \over 4^2} 
\langle\p X^{\eu} X^{\fu} (y) \p X^{\gu} X^{\hu}(z)\rangle
\d_{\eu}^{[\au}\d_{\fu}^{\bu ]}\d_{\gu}^{[\cu}\d_{\hu}^{\du ]}.$$ 
Each term can be computed either in momentum or coordinate space
without ambiguities. The first term gives 
\be
{\a^4 \over 4} \langle
\p X^\a X^{\bh} (y) \p X^\b X^{\ah}(z)\rangle (\g^{\au\bu})_\a {}^\g
\d_{\g\bh} (\g^{\cu\du})_\b {}^\d \d_{\d\ah} = - {\a^4 \over 4}{\eta^{\au [
\cu}\eta^{\du ]\bu}\over (y-z)^2}.
\ee
The second gives
\be 
- {\a^4 \over 4} \langle
\p X^\a X^{\bh} (y) \p X^{\ah} X^{\b}(z)\rangle (\g^{\au\bu})_\a {}^\g
\d_{\g\bh} (\g^{\cu\du})_{\ah} {}^{\dh} \d_{\b\dh} = - {\a^4 \over 4}{\eta^{\au [
\cu}\eta^{\du ]\bu} ln|y-z|^2\over (y-z)^2}.
\ee
The third gives the same result as the second and the fourth gives the
same result as the first. Finally, the fifth term gives
\be 
{\a^4 \over 4^2} 
\langle\p X^{\eu} X^{\fu} (y) \p X^{\gu} X^{\hu}(z)\rangle
\d_{\eu}^{[\au}\d_{\fu}^{\bu ]}\d_{\gu}^{[\cu}\d_{\hu}^{\du ]} = {\a^4
\over 8}{\eta^{\au [
\cu}\eta^{\du ]\bu} \over (y-z)^2} (1 + ln|y-z|^2).
\ee
Thus 
\be\label{J_0J_0R}
\langle J^{\au\bu} (y) J^{\cu\du} (z)\rangle = -{3\over 8}\a^4 {\eta^{\au [
\cu}\eta^{\du ]\bu} \over (y-z)^2} (1 + ln|y-z|^2).
\ee
One can easily check, given the vertices of (\ref{SXXXa}) that there
is no way to form one loop diagrams without classical current
contributions for $\langle J^{\au}(y) J^\b (z)\rangle $, 
$\langle J^{\au}(y) J^{\bh} (z)\rangle$, $\langle J^\a (y) J^\b
(z)\rangle$, $\langle J^{\ah} (y) J^{\bh} (z)\rangle$, $\langle
J^{\au} (y) J^{\bu\cu} (z)\rangle$, $\langle
J^{\a} (y) J^{\bu\cu} (z)\rangle$, $\langle
J^{\ah} (y) J^{\bu\cu} (z)\rangle$.

Let's compute now $\langle J^{\au}(y) \bar J^{\bu} (z)\rangle $
\be\label{J_2Jbar_2} \langle J^{\au} (y) \Jb^{\bu}(z)\rangle = \a^2 \langle
\p X^{\au}(y) \pb X^{\bu} (z) \rangle - \a^3 \langle \p X^{\au} (y) \pb
X^\a  X^{\b} (z)\rangle \g^{\bu}_{\a\b} - \a^3 \langle \p X^{\au} (y)
\pb X^{\ah}  X^{\bh} (z)\rangle \g^{\bu}_{\ah\bh}
\ee 
$$ - \a^3 \langle \p X^\a  X^{\b} (y) \pb X^{\bu}(z)\rangle \g^{\au}_{\a\b}
- \a^3 \langle \p X^{\ah}  X^{\bh} (y) \pb X^{\bu}(z)\rangle
\g^{\au}_{\ah\bh} + \a^4 \langle \p X^\a X^\b (y) \pb X^{\gh} X^{\dh} (z)
\rangle \g^{\au}_{\a\b} \g^{\bu}_{\gh\dh} $$ $$+ \a^4 \langle \p X^{\ah}
X^{\bh} (y) \pb X^{\g} X^{\d} (z)
\rangle \g^{\au}_{\ah\bh} \g^{\bu}_{\g\d}.$$
In this case the first term will give
\be\label{J_2Jbar_2I}
\a^2 \langle
\p X^{\au}(y) \pb X^{\bu} (z) \rangle = -{\a^4 \over 8}
\eta^{\au\cu}\eta^{\bu\du}(\g_{\cu})_{\a\b}(\g_{\du})_{\ah\bh}{k\over{|k|^2}}{\bar
k \over{|k|^2}}[\d^{\a\bh}\d^{\b\ah}\int d^d p {p\bar p \over{|p|^2}} {(k-p)(\bar k - \bar p
)\over{|k-p|^2}}  
\ee $$ - \d^{\a\ah}\d^{\b\bh} \int d^d p {p^2 \over {|p|^2}} {(k-p)^2
\over{|k-p|^2}} ]$$
$$  = - 2 \a^4 \eta^{\au \bu} \left[ 1 + \left({1\over \e} +1 - ln {|k|^2 \over
\mu^2 }\right)\right].$$

As in the case of $\langle J^{\au} (y) J^{\bu}(z)\rangle$ the second term
cancels with the third and the fourth with the fifth, i.e. the second
and third diagrams cancel independently. Nevertheless,
the sixth term gives
\be\label{J_2Jbar_2II}
\a^4 \langle \p X^\a X^\b (y) \pb X^{\gh} X^{\dh} (z)
\rangle \g^{\au}_{\a\b} \g^{\bu}_{\gh\dh} = - \a^4\eta^{\au\bu} \left[1
+ \left({1\over \e} +1 - ln {|k|^2 \over
\mu^2 }\right)\right],
\ee
and the senventh term in (\ref{J_2Jbar_2}) gives the same result. So,
differently from $J^{\au} (y) J^{\bu}(z)$ where the first and fourth
diagrams canceled, they add up for $\langle J^{\au} (y)
\Jb^{\bu}(z)\rangle$, giving 
\be\label{J_2Jbar_2RM}
\langle J^{\au} (y) \Jb^{\bu}(z)\rangle = - 4\a^4\eta^{\au\bu} \left[1 +
\left({1\over \e} +1 - ln {|k|^2 \over
\mu^2 }\right)\right],
\ee
which in coordinate space is 
\be\label{J_2Jbar_2RC}
\langle J^{\au} (y) \Jb^{\bu}(z)\rangle = 4 \a^4\eta^{\au\bu}
\left[\d^{(2)}(y,z)ln|y-z|^2 - {1\over |y-z|^2}\right],
\ee

In a completely analog way $\langle J^{\a}(y) \Jb^{\bh}(z) \rangle $ gives
\be\label{J_1Jbar_3RC}
\langle J^{\a} (y) \Jb^{\bh}(z)\rangle = {5\over 4} \a^4\d^{\a\bh}
\left[\d^{(2)}(y,z)ln|y-z|^2 - {1\over |y-z|^2}\right],
\ee
and
\be\label{JzeroJbzeroR}\langle J^{\au \bu} (y) \Jb^{\cu\du}(z)\rangle \to 
{3\over 8}\a^4 \eta^{\au
[\cu}\eta^{\du ] \bu} \left[\d^{(2)}(y,z)ln|y-z|^2  -
{1\over|y-z|^2}\right].
\ee
Summarizing, the only non-vanishing one-loop results are
(\ref{J_0J_0R}), (\ref{J_2Jbar_2RC}), (\ref{J_1Jbar_3RC}) and
(\ref{JzeroJbzeroR}), which are consistent with the results found in \cite{Mazzucato:2009fv}.

\section{OPEs of the Energy momentum tensor with the currents}

The energy momentum tensor is 
\be \label{EM}
T = -{1\over {\a^2}}  ({1\over 2 } J^{\au} J ^{\bu}
\eta_{\au \bu} -4\d_{\a\bh}  J^{\bh} J ^\a +
2N_{\au \bu}
J^{[\au \bu]} + 2\o_\a \partial \l^\a ),
\ee

\subsection{Tree level}
In this subsection we will compute $T(y) J^A (z)$ at tree level. Let's
start with $J^{\au}$. The result is

\begin{eqnarray} \label {TJ2}\langle T (y)
J^{\au} (z)\rangle  & =& 
{J^{\au}(z)\over{(y-z)^2}} + {1\over {y-z}}\left(\p J^{\au}(z) + [J_0
,J_2]^{\au}(z) - {1\over 2} [N , J_2]^{\cu} (z)\right) \\ \nonumber &&+ {(\bar y -
\bar z)\over{(y-z)^2}}\left(\pb J^{\au}(z)  + [\Jb_0 , J_2]^{\au}(z) - [J_3 ,
\Jb_3 ]^{\au}(z) - {1\over 2 } [N , \Jb_2]^{\au}(z) - {1\over 2 } [\Nh
, J_2]^{\au}(z)\right)
\end{eqnarray}
Note that the second line of the equations above vanishes by the use of the classical equations of motion, so there is no inconsistency from the fact that $\bar \partial T=0$. 
We will now explain how to arrive to this result.
From the first term in the energy momentum tensor we obtain
\be \label{T1J2}
-{1\over 2\a^2} \langle J^{\bu} J^{\cu} \eta_{\bu\cu} (y) J^{\au
}(z)\rangle = -
J^{\bu} \langle \p X^{\cu} \eta_{\bu\cu} (y) \partial X^{\au} (z)
\rangle - J^{\bu}
\langle [J,X]^{\cu} \eta_{\bu\cu} (y)\p X^{\au} (z) \rangle  -
J^{\bu}\langle \p X^{\cu}\eta_{\bu\cu}(y) [J,X]^{\au} (z) \rangle
\ee
Contracting using the propagator in the first term of the right hand side
we obtain the double pole, as well as the terms with $\p J^{\au}$ and $\pb
J^{\au}$ in (\ref{TJ2}). Now, the expansion of the action contains
terms of the form $\p X^{\au} \Jb^{\cu\du}X^{\eu}
\eta_{\eu[\cu}\d_{\du ]}^{\bu}\eta_{\au\bu}$ and $\pb X^{\au} J^{\cu\du}X^{\eu}
\eta_{\eu[\cu}\d_{\du ]}^{\bu}\eta_{\au\bu}$. Specifically, those
terms come from the expansion of $\eta_{\au\bu} J^{\au}\Jb^{\bu}$.
Those terms can contribute at tree level when contracting with the
first term in (\ref{T1J2}). The first gives the $[\Jb_0 ,
J_2]^{\au}$ in (\ref{TJ2}), while the second gives a $-[J_0 ,
J_2]^{\au}$ which exactly cancels with the second term in
(\ref{T1J2}). The third term in (\ref{T1J2}) gives the $[J_0 ,
J_2]^{\au}$ which appears in (\ref{TJ2}).

From the second term in the energy momentum tensor we obtain
\be \label{T2J2}
{4\over\a^2} \d_{\a\bh} \langle J^\a J^{\bh}(y) J^{\au} (z) \rangle =
4 \d_{\a\bh} J^\a \langle \p X^{\bh} (y) \p X^{\au} (z) \rangle -
4 \d_{\a\bh} J^{\bh} \langle \p X^\a (y) \p X^{\au} (z) \rangle +
\ee $$ 
4 \d_{\a\bh} J^\a \langle [J,X]^{\bh} (y) \p X^{\au} (z) \rangle -
4 \d_{\a\bh} J^{\bh} \langle [J,X]^{\a} (y) \p X^{\au} (z) \rangle +
4 \d_{\a\bh} J^\a \langle \p X^{\bh} (y) [J,X]^{\au} (z) \rangle $$ $$
-4 \d_{\a\bh} J^{\bh} \langle \p X^{\a} (y) [J,X]^{\au} (z) \rangle$$
Expanding $\d_{\a\bh}J^{\a}\Jb^{\bh}$ and $-3\d_{\a\bh}
J^{\bh}\Jb^{\a}$ in the action (\ref{adsaction}) we can form treel
level diagrams with the first term in (\ref{T2J2}) whose result
vanishes. Nevertheless, the tree level diagrams formed with those
expansions and the second term in (\ref{T2J2}) gives the $[J_3 ,
\Jb_3]$ in (\ref{TJ2}). The remaining terms in (\ref{T2J2}) vanish
because they give contributions of the form $J^{\a}J^\b \g^{\au}_{\a\b}$ or  
$J^{\ah}J^{\bh} \g^{\au}_{\ah\bh}$. 

From the third term in the energy momentum tensor we can easily
obtain $-[N,J]^{\au} (y-z)^{-1}$, while using the first term in
(\ref{T1J2}) and the expansion of $\Jb^{\au\bu}N_{\au\bu}$ in the
action (\ref{adsaction}) we obtain ${1\over2} [N,J]^{\au}(y-z)^{-1}$,
giving at the end the term $[N,J]^{\au}$ in (\ref{TJ2}). Similarly,
the first term in (\ref{T1J2}) contracted with the expansion of $\hat
N_{\au\bu} J^{\au\bu}$ gives the  $[\hat N,J]^{\au}$ in (\ref{TJ2}).
Finally, the last term in the energy momentum tensor contracted with
$J^{\au}$ will give a tree level contribution by forming tree-level
diagram contracting with $N^{(1)}_{\au\bu} \Jb^{\au} X^{\bu}$, which comes 
from the expansion of $N_{\au\bu}\Jb^{\au\bu}$. This contribution will
be the $[N,\Jb_2]^{\au}$ in (\ref{TJ2}). Note that using the
classical equations of motion, the second line in (\ref{TJ2})
vanishes. Then, classically $J^{\au}$ is not a primary field.

Similarly, we obtain 
\begin{equation} \label {TJ1}\langle T (y)
J^{\a} (z)\rangle   = 
{J^{\a}(z)\over{(y-z)^2}} + {1\over {y-z}}\left(\p J^{\a}(z) + [J_0
,J_1]^{\a}(z) - {1\over 2} [N , J_1]^{\a}(z)\right) 
\end{equation}
and an analog expression for $\langle T(y) J^{\ah} (z)\rangle$. Nevertheless, it can
be easily checked that at tree level, $T(y) J^{[\au\bu]}$ is regular. 

It is also interesting to know $\langle T(y) \Jb^{A}(z)\rangle$. Following the same
computation described in detail for $\langle T(y) J^{\au}\rangle$, we found

\be\label{TJb2} \langle T(y) \Jb^{\au} (z)\rangle = -J^{\au} (y) \d^{(2)} (y-z) - {[J_1
,\Jb_1]^{\au}(z) \over {y-z}} -{1\over 2}{[N , J_2]^{\au} (z)\over {\bar
y-\bar z}} -{1\over 2}{[N ,\Jb_2]^{\au} (z)\over {y- z}} + {1\over
2}{[\hat N ,J_2]^{\au}(z) \over {
y- z}} , 
\ee

\be\label{TJb1} \langle T(y) \Jb^{\a} (z)\rangle= -J^{\a} (y) \d^{(2)}
(y-z) +{1\over 2}{[N ,J_1]^{\a}(z) \over {\bar
y-\bar z}} +{1\over 2}{[N ,\Jb_1]^{\a}(z) \over {y- z}} - {1\over
2}{[\hat N ,J_2]^{\a}(z) \over {
y- z}} , 
\ee

\be\label{TJb0} \langle T(y) \Jb^{\au\bu}(z)\rangle = {[J_2
,\Jb_2]^{\au\bu}(z) \over {y-
z}} -{[J_1 ,\Jb_3]^{\au\bu}(z) \over {y- z}} - {[J_3 ,\Jb_1]^{\au\bu}(z) \over {
y- z}} , 
\ee

Note that these results are not inconsistent with $\bar\partial T=0$,
since, as usual, this derivative only gives contact terms in the right-hand 
side. 

\subsection{One-loop}
Next, we calculate the OPE´s between the energy momentum tensor and
the current $J^{\au}$  at one loop. We are going to show that there is no
contribution to this OPE. To this aim, we need to go up to one classical 
fied in the action and current expansions. In particular, we need to 
evaluate terms whith one classical field and three quantum fields in the action.

We are not going to show the details like in the last subsection and we just list the contribution of each diagram directly in coordinate space. 

The unique contribution  to $\langle T(y) J^{\underline{c}}(z)\rangle$  OPE
 come from $-{1\over{2\a^2}}\langle \eta_{\au\bu} J^{\au} J^{\bu} (y)
 J^{\underline{c}} (z)\rangle$ and ${4\over \a^2}\langle
 \delta_{\alpha\bh}  J^{\bh} J^{\a} J^{\underline{c}} (z)\rangle$. It
 will be shown now that these OPE's cancel separetly. Let us start
 with the first one. Expanding $J^{\au} J^{\au} (y)$ and $J^{\underline{c}} (z)$ up to one classical current, the expectation values we need to calculate come out as follow: 

 \begin{eqnarray}
-{1\over{2\a^2}} \left\langle J^{\au} J^{\au}(y) J^{\underline{c}}(z)
\right\rangle &=&
- \eta_{\au\bu} J^{\au} (y)\left\langle\partial X^{\bu} (y)
\partial X^{\cu} (z)  \right\rangle -\alpha \eta_{\au\bu}
\gamma^{\bu}_{\ah\bh} J^{\au} (y)\left\langle \partial X^{\ah}
X^{\bh} (y) \partial X^{\cu} (z) \right\rangle \nonumber\\
&-&\frac{\alpha^2}{4}\eta_{\au\bu}f^{\cu}_{[\gu\hu]\eu}f^{[\gu\hu]}_{\du\fu} 
J^{\du}(z) \left\langle \partial X^{\au}\partial X^{\bu}(y )X^{\eu} 
X^{\fu}(z) \right\rangle \\ \nonumber
&-&\frac{\alpha}{2} \eta_{\au\bu}\left\langle\partial X^{\au} 
 \partial X^{\bu} (y)\partial X^{\cu} (z) \right\rangle 
+\frac{\alpha^2}{4}\eta_{\au\bu}\gamma^{\cu}_{\underline{\a}\underline{\b}}
\left\langle\partial X^{\au}\partial X^{\bu}(y)\partial X^{\underline{\a}}
X^{\underline{\b}} (z) \right\rangle
 \nonumber \\
 &+&\a^2 \eta_{\au\bu} J^{\au}\left\langle \gamma^{\bu}_{\underline{\a}
 \underline{\b}}\partial X^{\underline{\a}} X^{\underline{\b}}(y)
 \gamma^{\cu}_{\underline{\a}\underline{\b}}\partial X^{\underline{\a}}
 X^{\underline{\b}}(z)\right\rangle\nonumber \\ \nonumber
 &+&\a \eta_{\au\bu}\gamma^{\cu}_{\underline{\a} \underline{\b}} J^{\bu}
 \left\langle\partial X^{\au} (y) \partial X^{\underline{\a}} 
 X^{\underline{\b}}(z) \right\rangle,
 \end{eqnarray}
where we are using the notation: $\underline{\a}\rightarrow (\a,\ah)$. For the sake of simplicity we don't write explicity the structure constants $f^{\du}_{[\au\bu]\cu}$ and $f^{[\au\bu]}_{\eu\fu}$. The first term is given by diagram five


\be \epsfbox{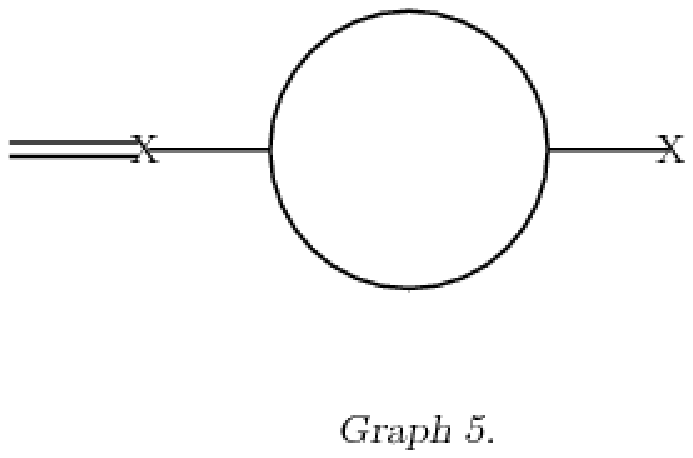} \label{picV}  \ee


The result is 
\begin{equation}
-\alpha^2\eta_{ab} J^a (y) \left\langle \partial X^b (y)\partial X^c
\right\rangle=  -\frac{2\alpha^2 J^{\underline{c}}(y)}{(y-z)^2}
(1+\ln |y-z|^2 )
\end{equation}
The next term is computed by evaluating diagram six


\be \epsfbox{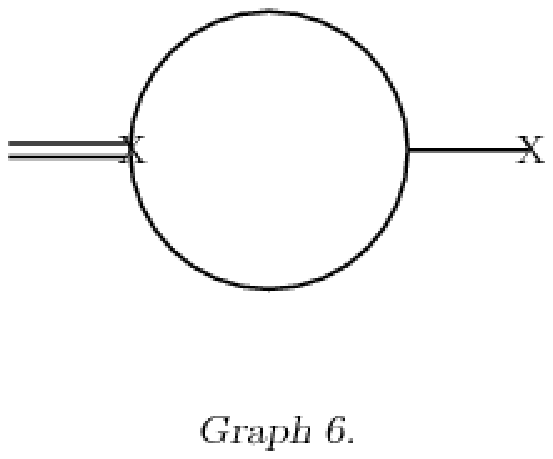} \label{picVI}  \ee


and the result is zero.  The contribution for the third term comes from diagram seven.


\be \epsfbox{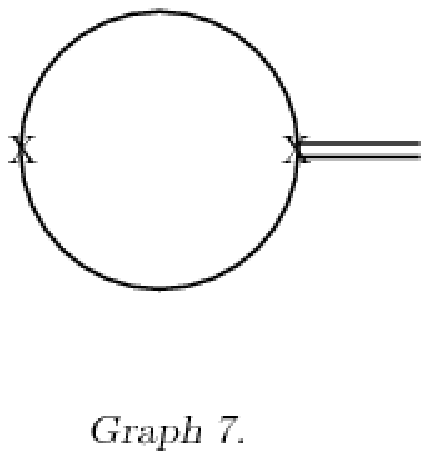} \label{picVI}  \ee


The result is
\be
-\frac{\alpha^4}{4}\eta_{\au\bu}f^{\cu}_{[\gu\hu]\eu}f^{[\gu\hu]}_{\du\fu}
J^{\du}(z) \left\langle \partial X^{\au}(y)\partial X^{\bu}(y )X^{\eu} X^{\fu}(z) \right\rangle = -\a^2 \frac{f^{\cu}_{[\au\bu]\fu}f^{[\au\bu]}_{\eu\fu} J^{\eu}(z)}{2(y-z)^2}
\ee

The next two terms could be calculated by evaluating diagrams eight and nine, but there are no possible contractions
and  they do not contribute.


\be \epsfbox{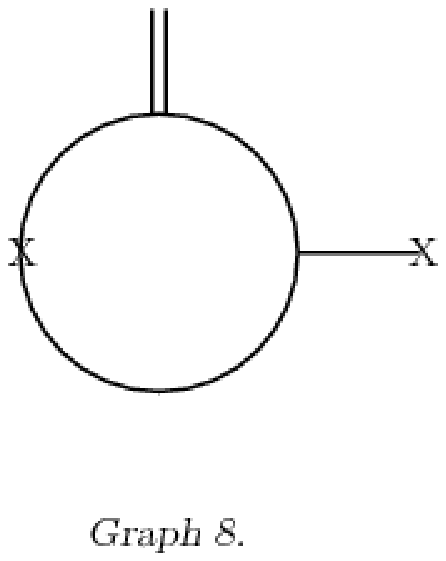} \label{picVI}  \ee


\be \epsfbox{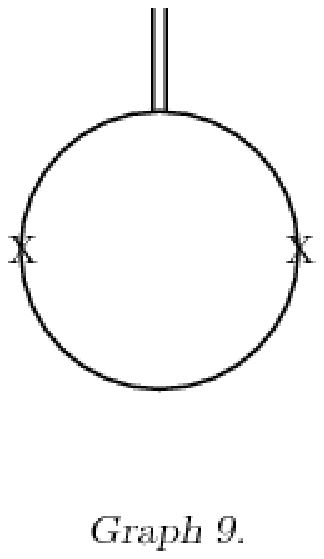} \label{picVI}  \ee


 The  contribution for the  sixth term comes from diagram ten and gives
\be \epsfbox{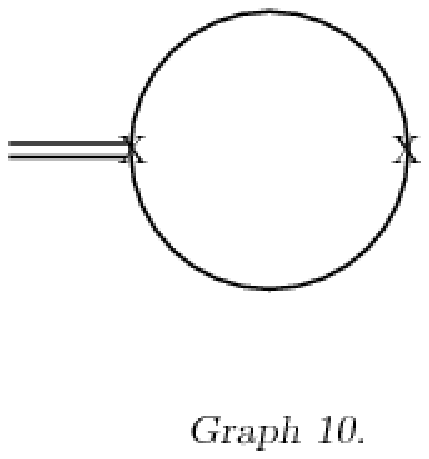} \label{picVI}  \ee

\be
+\a^2 \eta_{\au\bu} J^{\au} \left\langle \gamma^{\bu}_{\underline{\a}\underline{\b}}
\partial X^{\underline{\a}}(y) X^{\underline{\b}}(y)
\gamma^{\cu}_{\underline{\a} \underline{\b}} \partial
X^{\underline{\a}}(z) X^{\underline{\b}}(z)\right\rangle =\frac{2\alpha^2 J^{\underline{c}}(y)}{(y-z)^2}
(1+\ln |y-z|^2 ).
\ee
The seventh term is calculated from diagram eleven and the result is zero. 

\be \epsfbox{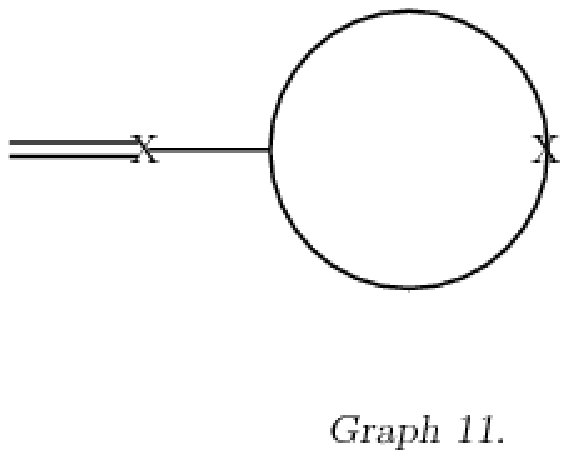} \label{picVI}  \ee


Finally, the fourth term also contributes to diagram twelve
\be \epsfbox{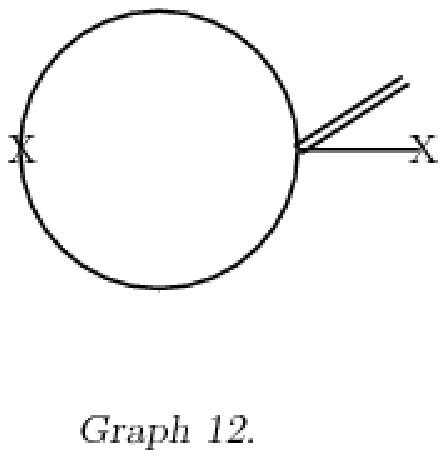} \label{picVI}  \ee
where it was used a vertex from the action which one classical field and three quantum fields. The result is
\be
-\frac{\alpha}{2}\eta_{\au\bu} \left\langle \partial X^{\au}(y)\partial X^{\bu}(y)\partial X^{\cu} \right\rangle
=\a^2\frac{f^{\cu}_{[\au\bu]\fu}f^{[\au\bu]}_{\eu\fu} J^{\eu}(z)}{2(y-z)^2} 
 \ee
So, we conclude that $-{1\over {2\a^2}}\left\langle \eta_{\au\bu}
J^{\au} J^{\bu} (y) J^{\underline{c}} (z)\right\rangle =0$.  

Now we will show
 that ${4\over \a^2}\left\langle \delta_{\alpha\bh}  J^{\bh} J^{\a} (y)J^{\underline{c}} (z)\right\rangle$ is also zero. Again we need to expand
the currents up to one classical field and calculate each expectation value. As the relevant diagrams are the same, 
we are not going to put the results for each expectation value and we will use  the notation $I_n$ for the n-th diagram, and just list the result of the diagrams that contribute, as follows 
\begin{eqnarray}
I_6&=& 0\nonumber \\
I_7&=&  -\frac{4\alpha^2 J^{\underline{c}}(z)}{(y-z)^2}
\nonumber \\
I_8 &=& \frac{2\alpha^2 J^{\underline{c}}(z) \ln
  |y-z|^2}{(y-z)^2} + \frac{3\alpha^2 J^{\underline{c}}(z)}{(y-z)^2}\nonumber\\
I_9 &=& \frac{2\alpha^2 J^{\underline{c}}(z)}{(y-z)^2}
(2+\ln |y-z|^2 )\nonumber\\
I_{10}&=&  -\frac{4\alpha^2 J^{\underline{c}}(y)}{(y-z)^2}
(1+\ln |y-z|^2 )\nonumber\\
I_{12} &=&  \frac{\alpha^2 J^{\underline{c}}(z)}{(y-z)^2} 
\end{eqnarray}

After evaluating the background fields at the point $(z, \bar{z})$, the sum 
of the diagrams is null. The derivative terms of the $J^{\underline{c}}$ 
don't appear in the results because they can be written as  bilinear terms 
in the classical fields due to the equations of motion, and they will
not enter in this one classical field calculation. Therefore, one can see 
that the result of the one loop calculation is

\begin{equation}
\langle T(y) J^{\underline{c}}(z) \rangle = 0
\end{equation}

Now, for the currents $J_1$ and $J_3$ the results are different. 
The one-loop results for $\langle -{1\over 2\a^2}\eta_{\au\bu} J^{\au}
J^{\bu} (y) J^\g (z) \rangle$ are

\begin{eqnarray}
I_6&=&  {5\over 4}\a^2 {J^\g (z) \over (y-z)^2} [1 + ln|y-z|^2],
\nonumber \\
I_7 &=& {5\over 4}\a^2 {J^\g (z) \over (y-z)^2},\nonumber\\
I_8 &=& -{5\over 4}\a^2 {J^\g (z) \over (y-z)^2} [{3\over 2} + ln|y
-z|^2],\nonumber\\
I_{9}&=&  - {5\over 4}\a^2 {J^\g (z) \over (y-z)^2} [2 + ln|y-z|^2],\nonumber\\
I_{10} &=&  {5\over 4}\a^2 {J^\g (z) \over (y-z)^2} [1 +
ln|y-z|^2],\nonumber \\
I_{12} &=& {5\over 8}\a^2 {J^\g (z) \over (y-z)^2},\nonumber\\
\end{eqnarray}
then in one-loop order $\langle -{1\over 2}\eta_{\au\bu} J^{\au}
J^{\bu} (y) J^\g (z) \rangle$ vanishes. Nevertheless, computing $\langle
-4\d_{\a\bh}J^{\a}J^{\bh} (y) J^\g (z) \rangle$ we found the following
results for each diagram

\begin{eqnarray}
I_5&=&  {5\over 4}\a^2 {J^\g (z) \over (y-z)^2} [1 + ln|y-z|^2],
\nonumber \\
I_6 &=& 0,\nonumber\\
I_7 &=& {5\over 4} \a^2 {J^{\g}(z)\over (y-z)^2} \nonumber \\
I_8 &=& -{5\over 4}\a^2 {J^\g (z) \over (y-z)^2} [{3\over 2} + ln|y
-z|^2],\nonumber\\
I_{9}&=&  - {5\over 4}\a^2 {J^\g (z) \over (y-z)^2} [2 + ln|y-z|^2],\nonumber\\
I_{10} &=&0,\nonumber \\
I_{11} &=&  {5\over 4}\a^2 {J^\g (z) \over (y-z)^2} [1 +
ln|y-z|^2],\nonumber \\
I_{12} &=& {5\over 16}\a^2 {J^\g (z) \over (y-z)^2},\nonumber\\
\end{eqnarray}
then $T(y) J_1 (z)$ does not cancel and indeed gives
\be \langle T(y) J^\g (z) \rangle =  -{5\over 16} \a^2 {J^\g (z) \over (y-z)^2}.
\ee

Something similar happens for $T(y) J_3 (z)$. Computing $-{1\over 2}
\eta_{\au\bu} \langle J^{\au} J^{\bu} (y) J^{\gh} (z)\rangle$ we found

\begin{eqnarray}
I_6 &=& -{5\over 4} \a^2 {J^{\gh}(z)\over (y-z)^2} [1 +
ln|y-z|^2] ,\nonumber \\
I_7 &=& {5\over 4} \a^2 {J^{\gh}(z) \over {(y-z)^2}}, \nonumber \\
I_8 &=& 0,\nonumber\\
I_{9}&=& 0, \nonumber \\
I_{10} &=& {5\over 4} \a^2 {J^{\gh}(z)\over (y-z)^2} [1 +
ln|y-z|^2],\nonumber \\
I_{12} &=& -{5\over 4}\a^2 {J^\g (z) \over (y-z)^2},\nonumber\\
\end{eqnarray}
so, $\langle -{1\over 2}\eta_{\au\bu} J^{\au} J^{\bu} (y) J^{\gh} (z)
\rangle$
cancels. Nevertheless the diagram results for $\langle -4 \d_{\a\bh}J^\a
J^{\bh} (y) J^{\gh} (z)\rangle$ are
\begin{eqnarray}
I_5 &=& {5\over 4} \a^2 {J^{\gh}(z)\over (y-z)^2} [1 +
ln|y-z|^2] ,\nonumber \\
I_6 &=& 0 , \nonumber\\
I_7 &=& {5\over 4} \a^2 {J^{\gh}(z) \over {(y-z)^2}}, \nonumber \\
I_8 &=& 0,\nonumber\\
I_{9}&=& 0, \nonumber \\
I_{10} &=& 0, \nonumber\\
I_{11} &=& -{5\over 4} \a^2 {J^{\gh}(z)\over (y-z)^2} [1 +
ln|y-z|^2],\nonumber \\
I_{12} &=& -{5\over 16}\a^2 {J^\g (z) \over (y-z)^2},\nonumber\\
\end{eqnarray}
So, 
\be \langle T(y) J^{\gh} (z) \rangle =  {5\over 16} \a^2 {J^{\gh} (z) \over (y-z)^2}.
\ee

Finally, let's consider $\langle T (y) J^{\cu\du} (z) \rangle$. Computing $\langle
-{1\over 2 \a^2}\eta_{\au\bu}J^{\au} J^{\bu}(y) J^{[cd]} (z)\rangle$ we
found
\begin{eqnarray}
I_7  &=&  \a^2 {J^{[cd]}(z)\over (y-z)^2}  ,\nonumber \\
I_9 &=&  - \a^2 {J^{[cd]}(z)\over (y-z)^2} [2 + ln |y-z|^2] , \nonumber\\
I_{10} &=&  \a^2 {J^{[cd]}(z)\over (y-z)^2} [1 + ln |y-z|^2], \nonumber \\
\end{eqnarray}
and the same result with opposite sign for $\langle
-{1\over 2\a^2}\eta_{\au\bu}J^{\au} J^{\bu}(y) J^{[c'd']} (z)\rangle$ so in one loop order $\langle
-{1\over 2\a^2}\eta_{\au\bu}J^{\au} J^{\bu}(y) J^{[\cu\du]} (z)\rangle$
cancels. 

Similarly, computing $\langle  -{4\over \a^2} \d_{\a\bh} J^{\a} J^{\bh} (y)
J^{cd} (z) \rangle$ we found

\begin{eqnarray}
I_7  &=&  4 \a^2 {J^{[cd]}(z)\over (y-z)^2}  ,\nonumber \\
I_9 &=&  - 4\a^2 {J^{[cd]}(z)\over (y-z)^2} [2 + ln |y-z|^2] , \nonumber\\
I_{10} &=&  4\a^2 {J^{[cd]}(z)\over (y-z)^2} [1 + ln |y-z|^2], \nonumber \\
\end{eqnarray}
and the same results with opposite sign for $\langle  -{4\over \a^2 } \d_{\a\bh} J^{\a} J^{\bh} (y)
J^{c'd'} (z) \rangle$. Then $\langle  -{4\over \a^2} \d_{\a\bh} J^{\a} J^{\bh} (y)
J^{\cu\du} (z) \rangle$ cancels at one loop order. Considering the
term $N_{\au\bu}J^{\au\bu}$ in the energy momentum tensor, we find
that the diagram 10 contributes
\be
I_{10} = {3\over 4}\a^2 {N^{\cu\du} \over {(y-z)^2 }}[1 + ln |y-z|^2],
\ee
and this result is directly related to (\ref{J_0J_0R}). This last
result is canceled by computing the one loop contribution coming from
the contraction of the last term in the energy momentum tensor
(\ref{EM}) with the term $N_{\au\bu}^{(1)} \Jb^{\au\bu}$ coming from
the expansion of the action . In conclusion, the one loop contribution
for $\langle T(y) J^{\cu\du} (z) \rangle$ cancels.

\section{Summary of Results}
In this work we showed that at one loop, there are non trivial cancelations 
in the possible corrections to the double pole of the product of the currents 
$J^{\au}(y) J^{\bu}(z)$ and $J^{\a}(y)
J^{\bh}(z)$. These results are in
agreement with \cite{Mazzucato:2009fv}. On the other hand, we
found the following one loop corrections to the double pole corrections
\be\label{J_0J_0RS}
\langle J^{\au\bu} (y) J^{\cu\du} (z)\rangle = -{3\over 8}\a^4 {\eta^{\au [
\cu}\eta^{\du ]\bu} \over (y-z)^2} (1 + ln|y-z|^2),
\ee
\be\label{J_2Jbar_2RCS}
\langle J^{\au} (y) \Jb^{\bu}(z)\rangle = -4 \a^4\eta^{\au\bu}
\left[{1\over |y-z|^2} -\d^{(2)}(y,z)ln|y-z|^2 \right] ,
\ee
\be\label{J_1Jbar_3RCS}
\langle J^{\a} (y) \Jb^{\bh}(z)\rangle = -{5\over 4} \a^4\d^{\a\bh}
\left[{1\over |y-z|^2}-\d^{(2)}(y,z)ln|y-z|^2 \right],
\ee
and
\be\label{JzeroJbzeroRS}\langle J^{\au \bu} (y) \Jb^{\cu\du}(z)\rangle
= 
-{3\over 8}\a^4 \eta^{\au
[\cu}\eta^{\du ] \bu} \left[{1\over|y-z|^2} - \d^{(2)}(y,z)ln|y-z|^2  \right].
\ee
We also found that there is no way to form one loop diagrams without classical
current contributions, i.e double pole corrections for $\langle J^{\au}(y) J^\b (z)\rangle $, 
$\langle J^{\au}(y) J^{\bh} (z)\rangle$, $\langle J^\a (y) J^\b
(z)\rangle$, $\langle J^{\ah} (y) J^{\bh} (z)\rangle$, $\langle
J^{\au} (y) J^{\bu\cu} (z)\rangle$, $\langle
J^{\a} (y) J^{\bu\cu} (z)\rangle$, $\langle
J^{\ah} (y) J^{\bu\cu} (z)\rangle$.

About the product of the energy momentum tensor with the currents we
found the following results on-shell
\begin{equation} \label {TJ2RS}\langle T (y)
J^{\au} (z)\rangle   =
{J^{\au}(z)\over{(y-z)^2}} + {1\over {y-z}}\left(\p J^{\au}(z) + [J_0
,J_2]^{\au}(z) - {1\over 2} [N , J_2]^{\cu} (z)\right),
\end{equation}
where we found a non trivial cancellation in the possible one-loop contribution to the
double pole. On the other hand, for the fermionic currents we found
\begin{equation} \label {TJ1RS}\langle T (y)
J^{\a} (z)\rangle   = 
(1-{5\over16}\a^2){J^\a (z)\over{(y-z)^2}} + {1\over {y-z}}\left(\p J^{\a}(z) + [J_0
,J_1]^{\a}(z) - {1\over 2} [N , J_1]^{\a}(z)\right), 
\end{equation}
\begin{equation} \label {TJ3RS}\langle T (y)
J^{\ah} (z)\rangle   = 
(1+{5\over16}\a^2){J^{\ah} (z)\over{(y-z)^2}} + {1\over {y-z}}\left(\p
J^{\ah}(z) + [J_0
,J_3]^{\ah}(z) - {1\over 2} [N , J_3]^{\ah}(z)\right).
\end{equation}
Thus, there are one loop corrections in the double poles. However,
forming a single operator $J^\a J^{\ah}$ those corrections cancels,
which means that the energy momentum tensor still has zero anomalous
dimension. It is worth to note that for $\langle T(y)
J^{\au\bu}(z)\rangle$
we found regular terms at tree level, while at one loop the possible
corrections to the double pole term cancel. In this cancelation plays
a key role the result (\ref{J_0J_0RS}) and the pure spinors. We also
computed $\langle T(y) \Jb^{A}(z) \rangle$ at tree level, whose
results were written at the send of subsection $6.1$.

In the one loop level of this work, we focused on the corrections to
the double poles. We leave the study of the possible corrections to
single poles for future work. 
\section*{Acknowledgements}

We would like to thank Nathan Berkovits, William Linch, Luca
Mazzucato, Andrei Mikhailov and Sakura Schafer-Nameki for useful
discussions. O.B would like to thank FAPESP grant 09/08893-9 and also 
the Instituto de Fisica Teorica at S\~ao Paulo for hospitality, as well as the Aspen Center of Physics
for hospitality during the workshop ``Unity of String Theory''. D. L. N. would like to  thank CNPq, grant 501317/2009-0, for
financial support.

\appendix
\section{Background Field Expansion}



Here we use the background field expansion described in section 4 and  write the expansion of the terms in the action (\ref{adsaction})
up to cubic terms in the quantum fields, since this is the
order relevant for the one-loop computation of the current's OPE's.
For the pure spinors Lorentz currents one expands
\be\label{pslcexp}N_{\au\bu} = N_{\au\bu}^{(0)} + \a N^{(1)}_{\au\bu} +
\a^2 N^{(2)}_{\au\bu},
\ee
and similarly for $\hat N_{\au\bu}$. Now, the pure spinor Lorentz
currents have the following behaviour

\be\label{NoneNone}N_{\au\bu}^{(1)} (y) N_{\cu\du}^{(1)}(z) \to 
{{\eta_{\cu[\bu} N_{\au]\du}^{(0)}(z) - \eta_{\du[\bu}
N_{\au]\cu}^{(0)}(z)}\over{y-z}},
\ee

\be\label{NtwoNtwo}N_{\au\bu}^{(2)} (y) N_{\cu\du}^{(2)}(z) \to
-3{{\eta_{\au[\du}\eta_{\cu]\bu}}\over{(y-z)^2}}.
\ee

\section{Explicit expansion of the action }
In this subsection we will write down the expansion of the matter part
of action containing three quantum fields and one classical current. 
 
The contributions proportional to $J^{\au}$ and $\Jb^{\au} $ are

\be\label{Jd}  {{\a^ 3 } \over 2 } \int d^2 z [{2\over 3}\p X^ a X^ b X^c \Jb^d
\eta_{a[b}\eta_{d]c} - {2\over 3} \p X^ {a'} X^ {b'} X^{c'} \Jb^{d'}
\eta_{a'[b'}\eta_{d']c'} +{2\over 3} \pb X^ a X^ b X^c J^d
\eta_{a[b}\eta_{d]c} 
\ee
$$- {2\over 3} \pb X^ {a'} X^ {b'} X^{c'} J^{d'}
\eta_{a'[b'}\eta_{d']c'}  -{1\over 3} \p X^{\au} X^\a
X^{\bh}\Jb^{\du}(\eta_{\au\du}\d_{\a\bh} + {5\over 4}
(\g_{\au\du})_{\bh}{}^{\gh} \d_{\a\gh} 
- {1\over 4} (\g_{\au\du})_\a {}^\g
\d_{\g\bh})$$ $$
-{1\over 3} \pb X^{\au} X^\a
X^{\bh}J^{\du}(\eta_{\au\du}\d_{\a\bh} - {1\over 4}
(\g_{\au\du})_{\bh}{}^{\gh} \d_{\a\gh} 
+ {5\over 4} (\g_{\au\du})_\a {}^\g
\d_{\g\bh}) -{1\over 3} \p X^\a X^{\bh} (X^{\au}\Jb^{\du}
\eta_{\au\du}\d_{\a\bh} +{1\over 2} X^a \Jb^d
(\g_{ad})_\a {}^{\gh}\d_{\b\gh}$$ $$
+{3\over 2} X^{a'} \Jb^{d'}
(\g_{a' d' })_\a {}^{\gh}\d_{\b\gh}) + {2\over 3} \pb X^\a X^{\bh} 
(X^{\au}J^{\du}\eta_{\au\du}\d_{\a\bh} +{11\over 4} X^a J^d
(\g_{ad})_\a {}^{\g}\d_{\g\bh}$$ $$
-{3\over 4} X^{a'} J^{d'}
(\g_{a' d'})_{\a}{}^{\g}\d_{\g\bh}) -{2\over 3} \p X^{\bh} X^\a
(X^{\au}\Jb^{\du}\eta_{\au\du}\d_{\a\bh} + {11\over 4}X^a \Jb^d
(\g_{ad})_{\bh}{}^{\gh}\d_{\a\gh}$$ $$
-{3\over 4}X^{a'} \Jb^{d'}
(\g_{a' d'})_{\bh}{}^{\gh}\d_{\a\gh} +{1\over 3} \pb X^{\bh} X^\a
(X^{\au}J^{\du}\eta_{\au\du}\d_{\a\bh} + {1\over 2} X^a J^d
(\g_{ad})_\a {}^{\g}\d_{\g\bh} + {3\over 2} X^{a'}J^{d'} (\g_{a'
d'})_{\a}{}^\g \d_{\g\bh})].$$ 

The contributions proportional to $J^\d$ are

\be\label{Jbardelta}\a^3  \int d^2 z [{2\over 3} \pb X^\a X^\b X^\g
(\g^{\au})_{\d\g}(\g_{\au})_{\a\b} -{1\over 3} \pb X^{\ah}X^{\bh} X^\g 
((\g^{\au})_{\g\d} (\g_{\a})_{\ah\bh}  + (\g^{ab})_\d {}^\e
\d_{\e\bh}(\g_{ab})_{\ah}{}^{\eh}\d_{\g\eh} -(\g^{a'b'})_\d {}^\e
\d_{\e\bh}(\g_{a'b'})_{\ah}{}^{\eh}\d_{\g\eh} )
\ee
 $$-{5\over 24} \pb
X^{\ah} X^{\au}X^{\bu}\d_{\d\ah}\eta_{\au\bu} - {5\over 6}\pb X^\a
X^{\bh} X^{\gh}((\g^{ab})_\a {}^\e \d_{\e\bh}(\g_{ab})_\d {}^\b
\d_{\b\gh} - (\g^{a'b'})_\a {}^\e \d_{\e\bh}(\g_{a'b'})_\d {}^\b
\d_{\b\gh}  )
$$ $$ + {1\over 12} \pb X^{\au} X^{\bu} X^{\ah}\d_{\d\ah}\eta_{\au\bu}
 + {1\over 4} \pb X^a X^b X^{\ah}\d_{\d\bh}(\g_{ab})^{\bh}{}_{\ah} +
 {13\over 12} \pb X^{a'} X^{b'}
 X^{\ah}\d_{\d\bh}(\g_{a'b'})^{\bh}{}_{\ah}]J^\d .$$

The contributions propotional to $\Jb^{\dh}$ are 
\be\label{Jbardelhat}\a^3 \int d^2 z [{2\over 3} \p X^{\ah} X^{\bh} X^{\gh}
(\g^{\au})_{\dh\gh}(\g_{\au})_{\ah\bh} -{1\over 3} \p X^{\a}X^{\b}
X^{\gh}
((\g^{\au})_{\gh\dh} (\g_{\a})_{\a\b}  + (\g^{ab})_{\dh} {}^{\eh}
\d_{\b\eh}(\g_{ab})_{\a}{}^{\e}\d_{\e\gh} -(\g^{a'b'})_{\dh} {}^{\eh}
\d_{\b\eh}(\g_{a'b'})_{\a}{}^{\e}\d_{\e\gh} )
\ee
 $$+{5\over 24} \p
X^{\a} X^{\au}X^{\bu}\d_{\a\dh}\eta_{\au\bu} - {5\over 6}\p X^{\ah}
X^{\b} X^{\g}((\g^{ab})_{\ah} {}^{\eh} \d_{\b\eh}(\g_{ab})_{\dh}
{}^{\bh}
\d_{\g\bh} - (\g^{a'b'})_{\ah} {}^{\eh} \d_{\b\eh}(\g_{a'b'})_{\dh}
{}^{\bh}
\d_{\g\bh}  )
$$ $$ - {1\over 12} \p X^{\au} X^{\bu} X^{\a}\d_{\a\dh}\eta_{\au\bu}
 - {1\over 4} \p X^a X^b X^{\a}\d_{\b\dh}(\g_{ab})^{\b}{}_{\a} -
 {13\over 12} \p X^{a'} X^{b'} X^{\a}\d_{\b\dh}(\g_{a'b'})^{\b}{}_{\a}
]\Jb^{\dh}. $$



\section{List of integrals}

\be\label{integralzero} \int d^d m  {{1}\over{|m|^2|m-k|^2}} = - {2\over {k
\bar k}} ({1\over \epsilon} - ln {|k|^2 \over \mu^2}).
\ee

\be\label{integralI}\int d^d m  {{m \bar m}\over{|m|^2|m-k|^2}} = 1.
\ee

\be\label{integralII}\int d^d m {m\over{|m|^2 |m-k|^2}} = -{1\over \pi^\e} 
{1\over \bar k}{{[|k|^2 ]^{-\e}}\over{\mu^{-2\e}}} {{\G (1-\e)^2
\G(\e)}\over{\G(1-2\e)}} = - {1\over{\bar k}} ({1\over \e} - ln{|k|^2
\over{\mu^2}}).
\ee

\be\label{integralIII}\int d^d  m {{\bar m}\over{|m|^2 |m-k|^2}} = -{1\over \pi^\e} 
{1\over k}{{[|k|^2 ]^{-\e}}\over{\mu^{-2\e}}} {{\G (1-\e)^2
\G(\e)}\over{\G(1-2\e)}} = - {1\over{ k}} ({1\over \e} - ln{|k|^2
\over{\mu^2}}).
\ee

\be\label{integralIV}\int d^d m {m^2\over{|m|^2 |m-k|^2}} = -{1\over \pi^\e} 
{k\over \bar k}{{[|k|^2 ]^{-\e}}\over{\mu^{-2\e}}} {{\G(2-\e)\G (1-\e)
\G(\e)}\over{\G(2-2\e)}} = - {k\over \bar k} ({1\over \e} + 1 -
ln{|k|^2 \over \mu^ 2}).
\ee

\be\label{integralV}\int d^d m {{\bar m}^2\over{|m|^2 |m-k|^2}} = -{1\over \pi^\e} 
{\bar k\over k}{{[|k|^2 ]^{-\e}}\over{\mu^{-2\e}}} {{\G(2-\e)\G (1-\e)
\G(\e)}\over{\G(2-2\e)}} = - {\bar k\over k} ({1\over \e} + 1 -
ln{|k|^2 \over \mu^ 2}).
\ee

\be\label{integralVI}\int d^d m {m^2 {\bar m}\over{|m|^2 |m-k|^2}} =
{1\over 2\pi^\e} 
k{{[|k|^2 ]^{-\e}}\over{\mu^{-2\e}}} {{\G(1-\e) ^2 \G (1+\e)
}\over{\G(2-2\e)}} = {k\over 2}.
\ee

\be\label{integralVII}\int d^d m {m {\bar m}^2\over{|m|^2 |m-k|^2}} =
{1\over 2\pi^\e} 
{\bar k}{{[|k|^2 ]^{-\e}}\over{\mu^{-2\e}}} {{\G(1-\e) ^2 \G (1+\e)
}\over{\G(2-2\e)}} = {{\bar k}\over 2}.
\ee

\be\label{integralVIII}\int d^d m {m^2 {\bar m}^2\over{|m|^2 |m-k|^2}} =
{1\over \pi^\e} 
k{\bar k}{{[|k|^2 ]^{-\e}}\over{\mu^{-2\e}}} {{\G(2-\e) ^2 \G (1+\e)
}\over{\G(4-2\e)}} = {{k\bar k}\over 6}.
\ee

\be\label{integralIX}\int d^d m {m {\bar m}\over{[|m|^2]^2 |m-k|^2}} = -
{1\over \pi^\e} 
{1\over{k\bar k}}(1+ {2\over \epsilon} - 2 ln {|k|^2 \over \mu^2 } ).
\ee

\be\label{integralX}\int d^d m {m^2 {\bar m}\over{[|m|^2]^2 |m-k|^2}} = -
{1\over \pi^\e} 
{1\over{\bar k}}({1\over \epsilon} -  ln {|k|^2 \over \mu^2 } )  .
\ee

\be\label{integralXI}\int d^d m {m^3 {\bar m}\over{[|m|^2]^2 |m-k|^2}} = -
{1\over \pi^\e} 
{k\over{\bar k}}({1\over \epsilon} +1 -  ln {|k|^2 \over \mu^2 } ).
\ee

\be\label{integralXII}\int d^d m {m^2 {\bar m^2}\over{[|m|^2]^2
|m-k|^2}} = {3\over 2}.
\ee

\end{document}